\documentclass[letterpaper, numberedappendix]{emulateapj} 

\usepackage{graphicx}
\usepackage{epsfig}
\usepackage{amsmath}
\usepackage{subfigure} 
\usepackage{hyperref}
\usepackage{amssymb}
\usepackage{lscape} 
\usepackage{apjfonts}
\usepackage{latexsym}

\shorttitle{VV304a corrugation pattern}
\shortauthors{G\'omez et al.}

\begin{document}


\title[]{A tidally induced global corrugation pattern in an external disc galaxy similar to the Milky Way}


\author{Facundo A. G{\'o}mez\altaffilmark{1,2}}
\author{Sergio Torres-Flores\altaffilmark{1,2}}
\author{Catalina Mora-Urrejola\altaffilmark{2}} \author{Antonela Monachesi\altaffilmark{1,2}} 
\author{Simon D. M. White\altaffilmark{3}}
\author{Nicolas P. Maffione\altaffilmark{4,5}}
\author{Robert J. J. Grand\altaffilmark{3}} 
\author{Federico Marinacci\altaffilmark{6}} 
\author{{R\"u}diger Pakmor\altaffilmark{3}} 
\author{Volker Springel\altaffilmark{3}}
\author{Carlos S. Frenk\altaffilmark{7}} 
\author{Philippe Amram\altaffilmark{8}}
\author{Beno{\^i}t Epinat\altaffilmark{8}} 
\author{Claudia Mendes de Oliveira\altaffilmark{9}}

\altaffiltext{1}{Instituto de Investigaci\'on Multidisciplinar en Ciencia y Tecnolog\'ia, Universidad de La Serena, Ra\'ul Bitr\'an 1305, La Serena, Chile}
\altaffiltext{2}{Departamento de F\'isica y Astronom\'ia, Universidad de La Serena, Av. Juan Cisternas 1200 Norte, La Serena, Chile}
\altaffiltext{3}{Max-Planck-Institut f\"ur Astrophysik, Karl-Schwarzschild-Str. 1, 85748 Garching, Germany}
\altaffiltext{4}{Universidad Nacional de R\'{i}o Negro. R\'{i}o Negro, Argentina}
\altaffiltext{5}{Consejo Nacional de Investigaciones Cient\'{i}ficas y T\'{e}cnicas. Argentina}
\altaffiltext{6}{Department of Physics \& Astronomy, University of Bologna, via Gobetti 93/2, 40129 Bologna, Italy}
\altaffiltext{7}{Institute for Computational Cosmology, Department of Physics, University of Durham, South Road, Durham DH1 3LE, UK}
\altaffiltext{8}{Aix Marseille Univ, CNRS, CNES, LAM, Marseille, France}
\altaffiltext{9}{Departamento de Astronomia, Instituto de Astronomia, Geof{\'i}sica e Ci{\^e}ncias Atmosf{\'e}ricas da USP, Cidade Universitaria, CEP: 05508-900, S{\~a}o Paulo, SP, Brazil}

\begin{abstract}

We study the two dimensional (2D) line-of-sight velocity ($V_{\rm los}$) field  of the low-inclination, late-type galaxy 
VV304a. The resulting 2D kinematic map reveals a global, coherent and extended perturbation that is likely associated with a recent interaction with the massive 
companion VV304b. We use multi-band imaging and a suite of test particle simulations to quantify the plausible strength of in-plane flows due to non-axisymmetric perturbations and show that the observed velocity flows are much too large to be driven either by spiral structure nor by a bar. We use fully cosmological hydrodynamical simulations to characterize the contribution from in- and off-plane velocity flows to the $V_{\rm los}$ field of 
recently interacting galaxy pairs like the VV304 system. We show that, for recently perturbed low-inclination galactic discs, the structure of the residual velocity field,  after subtraction of an axisymmetric rotation model, can be dominated by vertical flows. Our results indicate that the  $V_{\rm los}$ perturbations in VV304a are consistent with a corrugation pattern. Its $V_{\rm los}$ map suggests the presence of a structure similar to the Monoceros ring seen in the Milky Way. 
Our study highlights the possibility of addressing important questions regarding the nature and origin of vertical perturbations by 
measuring the line-of-sight velocities in low-inclination nearby galaxies.

\end{abstract}





\section{Introduction}
\label{sec:intro}

Over the last decade a significant number of observational studies have provided strong evidence of an oscillating vertical asymmetry in our own Galactic disc \citep[e.g.][]{2012ApJ...750L..41W,2015ApJ...801..105X}. Well-known features observed in the Milky Way, such as the Monoceros Ring \citep{new02,yanny03}, the TriAnd clouds \citep{2015MNRAS.452..676P}, and the recently discovered A13 overdensity \citep{2018ApJ...854...47S}, can be naturally accounted for by this asymmetry, best described as a corrugation pattern \citep{2018MNRAS.481..286L}. With the recent arrival of the {\it Gaia} data release 2 \citep{2018arXiv180409380G}, we have had direct confirmation that our own Galactic disc is undergoing phase mixing of a non-equilibrium perturbation \citep{2018Natur.561..360A}. Most successful models to describe the origin of this perturbation  rely on the interaction between the Galaxy and the Sagittarius dwarf spheroidal galaxy, together with a recent boost from the Magellanic Clouds system \citep[e.g.][]{2012MNRAS.423.3727G,2013MNRAS.429..159G,2018MNRAS.473.1218L,2018MNRAS.481..286L,2019MNRAS.485.3134L,2018arXiv180902658B}. 

A corrugation pattern is manifested by an extended and oscillatory vertical displacement of the disc with respect to the overall disc mid-plane. Studies that have assessed the frequency with which corrugation patterns arise in late-type galaxies, based on cosmological simulations, have shown that such patterns are expected to be common \citep{2016MNRAS.456.2779G,2017MNRAS.465.3446G}. Large observational samples of nearly edge-on disc galaxies have revealed that $\sim$ 70 per cent show perturbed vertical structure \citep[e.g.][]{1998A&A...337....9R,2006NewA...11..293A}, typically displaying S-shaped warps. However, evidence for more complex corrugation patterns in external galaxies is still extremely limited \citep{1998AstL...24..764F,2001ApJ...550..253A,2015MNRAS.454.3376S}, and no full two dimensional (2D) maps of coherent and global features, similar to those observed in the Milky Way, have been reported. Morphological signatures of these vertical perturbations in external galaxies are washed out due to projection effects, independently of the inclination of the galactic disc. Thus, they cannot be detected through classical  imaging. Instead, thanks to the expected $90^{\circ}$ difference in oscillation phase between the local disc's mean-height, $\langle Z \rangle$, and mean vertical velocity, $\langle V_{\rm Z} \rangle$ \citep[e.g.][]{2013MNRAS.429..159G}, a corrugation pattern  can be revealed through line-of-sight velocity fields, $V_{\rm los}$, of low-inclination discs \citep[e.g.][]{2017MNRAS.465.3446G}.  In traditional long-lit spectroscopic observations, such patterns can be easily confused with the effects of local perturbations such as fountain flows \citep{2015MNRAS.454.3376S}. Therefore, full 2D kinematic maps are preferable to search for global and coherent patterns. Low inclination angles are also required to avoid significant contamination from in-plane flows in the resulting  $V_{\rm los}$ maps \citep[e.g.][]{2016MNRAS.460L..94G}.

Here, we present a full 2D kinematic map of a Milky Way-type galaxy, namely VV304a,  that shows a global, coherent and extended perturbation consistent with a corrugation pattern, likely associated with the recent interaction with its massive companion, VV304b.

\begin{figure}
    \centering
    \includegraphics[width=80mm,clip]{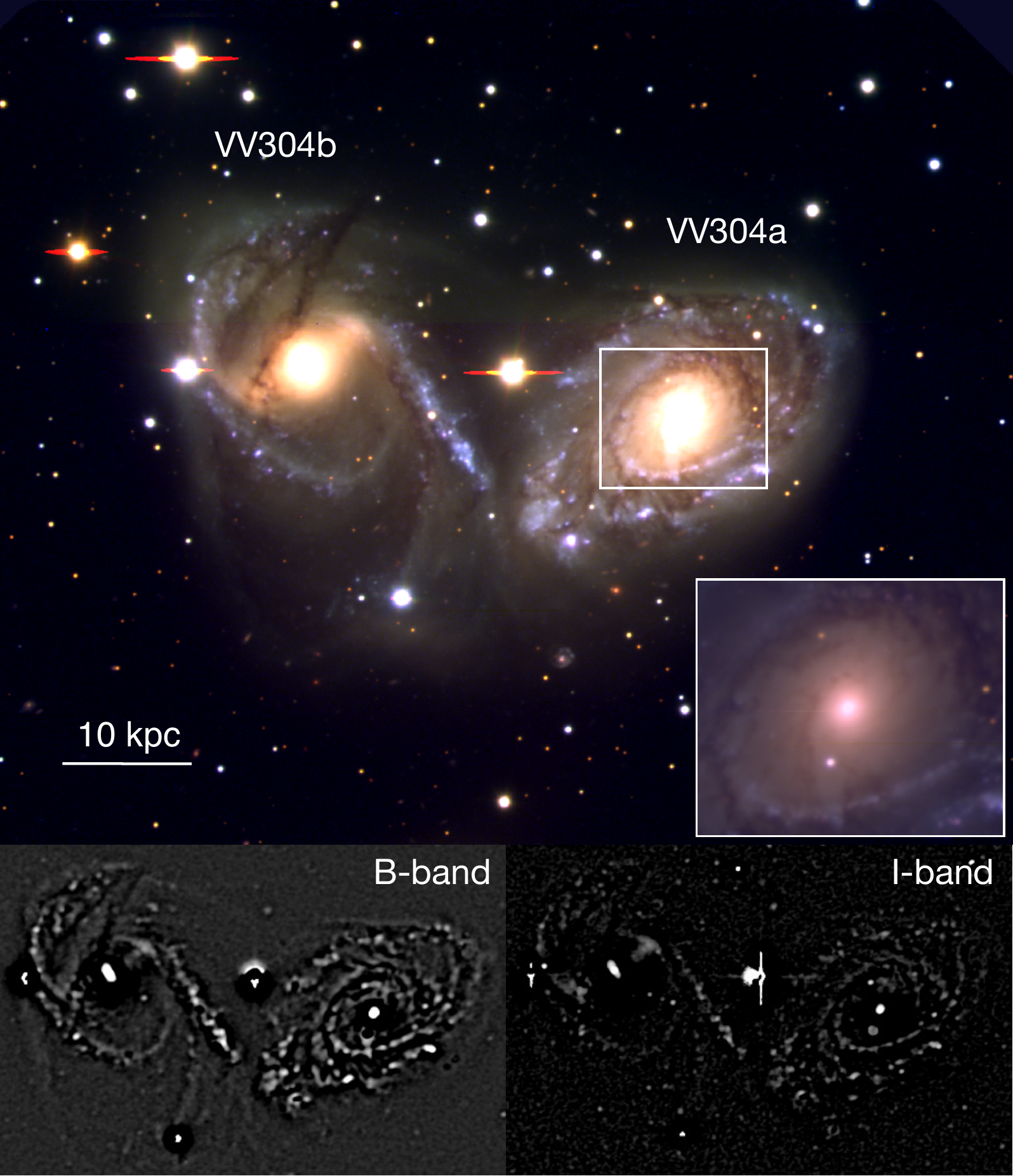}
    \caption{ Top panel: Optical image of the interacting pair VV304.  Both galaxies show clear indications of recent interaction. The u', g' and r'-band false colour image was obtained with the GMOS instrument at the Gemini South Observatory, under the scientific program GS-2013B-Q-27. Blue, green and red colours are associated with each filter, respectively. The inset (white square) allows to visually inspect inner galactic regions with significant more detail. Note that, unlike VV304b, VV304a does not posses a well-defined bar. Bottom panels: The left and right panels show structure maps in the B- and I-band, respectively.
    The maps were obtained from the publicly available database of the \href{https://cgs.obs.carnegiescience.edu/CGS/Home.html}{The Carnegie-Irvine Galaxy Survey}, and presented by \citet{2011ApJS..197...21H}.  Structure maps, originally developed by \citet{2002ApJ...569..624P}, are a powerful method to enhance fine structural features in single-filter images. The structure maps obtained in both filters clearly show the absence of a strong bar in the galaxy VV304a. In addition, they also highlight the lack of a dominant bisymmetric spiral structure. Interestingly, one can clearly observe the double strong barred nature of VV304b, demonstrating the strength of this method.}
    \label{fig:1}
\end{figure}

\section{The perturbed velocity field of VV304a}

\subsection{VV304a: General properties}
\label{sec:2-1}

VV304a is a late-type galaxy that has recently interacted with its closest companion, VV304b \citep[][hereafter TF14]{2014MNRAS.442.2188T}. VV304a is located at a distance of $\sim 54$ Mpc, has an optical radius of $R_{\rm opt} \approx 18$ kpc and a rotational velocity curve that peaks at $\sim 245$ km s$^{-1}$ (TF14). Based on the observed circular velocity profile, obtained using $H_{\alpha}$ Fabry-Perot observations (see Sec~\ref{sec:2.2}), TF14 estimated a total VV304a dynamical mass within $R_{\rm opt}$ of $M (<R_{\rm opt}) \approx 2.5 \times 10^{11}$ M$_{\odot}$. These broad properties are similar to those of the Milky Way \citep{2017MNRAS.465...76M, 2017RAA....17...96L, 2020MNRAS.494.4291C}. Its companion, VV304b, also a late-type galaxy, has an optical radius of 18 kpc and  a peak circular rotational velocity of $\sim 200$ km s$^{-1}$. The top panel of Figure~\ref{fig:1} shows a $u'$, $g'$ and $r'$-band Gemini/GMOS false colour image of the VV304 pair. The pair shows clear signatures of recent interaction, which makes VV304a an ideal candidate to search for a corrugated vertical structure. Visual inspection of this panel reveals that, unlike VV304b which shows a clear double barred structure and well defined grand design  spiral arms, VV304a does not possess any clear strong $m=2$ perturbation.  This is  particularly relevant for our study since strong $m=2$ modes could drive large in-plane velocity flows that could contribute significantly to, or even dominate, perturbations in the $V_{\rm los}$ field of VV304a.

\begin{figure}
    \centering
    \includegraphics[width=85mm,clip]{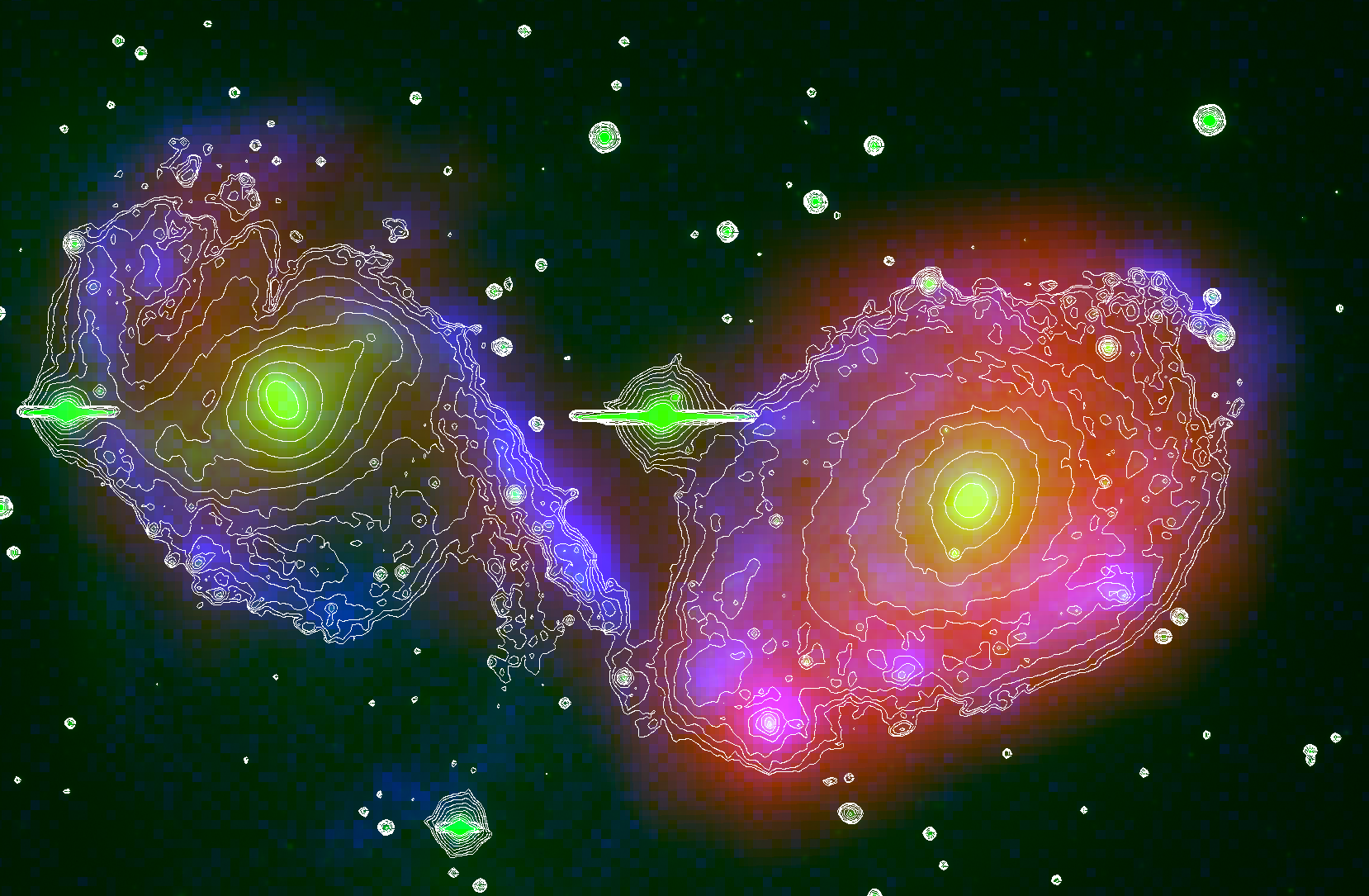}
    \caption{Mulitband image of the VV304 galactic pair, with isophotal contours. This image contains observations from Wise 22 microns (red), FUV GALEX (1530 \AA, Blue) and the r'-band optical image (Gemini/GMOS, green). The isophotes were derived from the r'-band image.}
    \label{fig:2}
\end{figure}

To better characterize the morphological structure of VV304a we show, on the left and right bottom panels of Fig.~\ref{fig:1}, structure maps in the B- and I-band, respectively. Structure maps, originally developed by \citet{2002ApJ...569..624P}, are a powerful method to enhance fine structural features in single-filter images. The maps were obtained from the publicly available database of the \href{https://cgs.obs.carnegiescience.edu/CGS/Home.html}{The Carnegie-Irvine Galaxy Survey}, and presented by \citet{2011ApJS..197...21H}. The structure maps obtained in both filters clearly show the absence of a  strong bar in VV304a. In addition, they also highlight the lack of a dominant bisymmetric spiral structure. Interestingly, one can clearly see the strongly double barred nature of VV304b, demonstrating the strength of this method.

In Figure~\ref{fig:2} we show a multiband image of the VV304 galactic pair. To create this image we have
used observations from:
\begin{itemize}
    \item Red: Wise 22 microns, (heated dust by UV photons).
    \item Blue: FUV GALEX (1530 \AA), UV photons associated with massive (young) stars.
    \item Green: r'-band optical image (Gemini/GMOS). Reddest optical band we have available.
\end{itemize}

The lack of any strong bar in the inner regions of VV304a is also evident from this image.
The iso-contour curves highlight the very round shape of the inner regions, only distorted by
the presence of a foreground star. In the outer regions, the iso-contours are distorted, reflecting the recent tidal interaction with VV304b.

\begin{figure*}
    \centering
    \includegraphics[width=180mm,clip,angle=0]{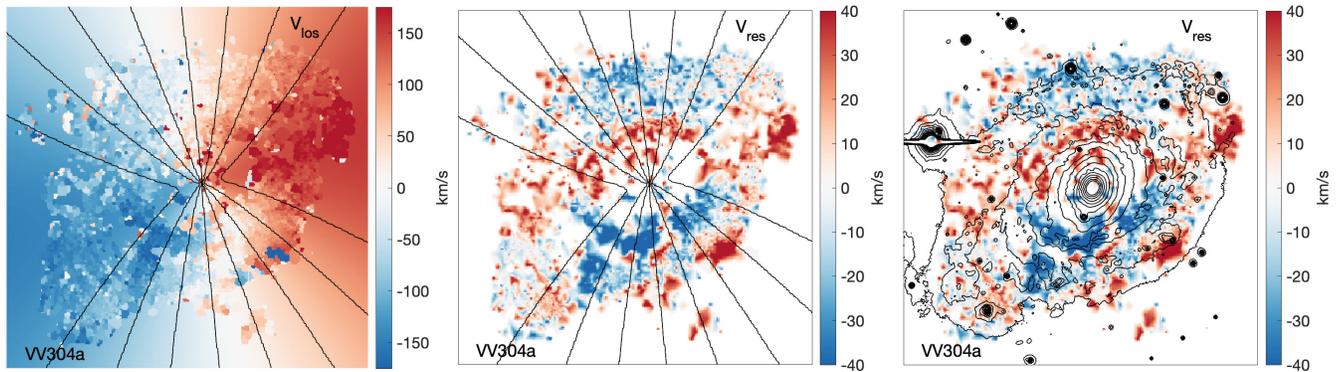}
    \caption{Left panel: Observed line-of-sight velocity field, $V_{\rm los}$, of VV304a, derived from its H$_{\alpha}$ emission. The smooth coloured background shows the modeled axisymmetric velocity field in regions without observable H$_{\alpha}$. The black lines show iso-velocity contours of this axisymmetric field. Middle panel: Residual velocity field, $V_{\rm res}$, obtained from the difference between the observed and modeled $V_{\rm los}$ field. The transitions from positive to negative velocities along a given iso-velocity line are consistent with a corrugation pattern. These velocity fluctuations could  indicate a significant displacement of the disc's $\langle Z \rangle$ with respect to its overall mid-plane. Right panel: Residual velocity field including isophotal contours obtained from r'-band optical image (Gemini/GMOS).}
    \label{fig:3}
\end{figure*}

\subsection{$H_{\alpha}$ Fabry-Perot observations}
\label{sec:2.2}

In this work we analyze the 2D kinematic field of VV304a's $H_{\alpha}$ distribution, obtained from Fabry-Perot observations published by TF14. The data were obtained with the instrument CIn{\'e}matique des GALaxiEs (CIGALE) mounted on the European Southern Observatory (ESO) 3.6 m telescope at La Silla (Chile). These types of observations are ideal to derive 2D kinematic maps due to their large field-of-view ($207 \times 207$ arcsec$^{2}$), high spectral resolution ($R \sim 13000$) and signal to noise ratios (S/N$\sim 5$). The data reduction process has been extensively described in TF14. 
Here we summarize the main steps taken to derive the two dimensional velocity fields.


The $H_{\alpha}$ 3D data cube of VV304a was reduced using a publicly available package, which is described in detail in  \citet{2006MNRAS.368.1016D}\footnote{The code is available at http://www.astro.umontreal.ca/fantomm/reduction/}. The package provides the velocity moment maps of the $H_{\alpha}$ 3D data cube and it has been used by different authors during the last years \citep[e.g.][]{2018A&A...614A..56B,2018MNRAS.480.3257M,2019A&A...631A..71G}. The main advantage of the techniques implemented in this package is the use of adaptive spatial binning, based on a 2D Voronoi tessellation method, applied to the spatial dimensions of the 3D data cubes. As a result, it allows one to obtain high spatial resolution in high signal-to-noise regions and large spatial coverage in low signal-to-noise regions. Here we demanded a uniform Signal-to-Noise ratio (SNR) over the whole field of view. This was achieved by binning neighboring pixels, until a signal-to-noise target (SNRt) is reached. For each bin, the noise is determined from the ratio between the $H_{\alpha}$ line flux and the root mean square of the continuum. Note that  regions of initial SNR higher than SNRt are not binned, thus keeping the angular resolution as high as possible. In this work, SNRt = 6 was used. To remove the OH sky lines, the reduction process of the Fabry-Perot data produces a sky-cube, which is removed from the observed cube. In practice, the sky cube is derived from sky dominated regions as follows. Considering that the $H_{\alpha}$ emission line of a galaxy displays small shifts in frequency due to the rotation of the gas, the sky emission lines correspond to the ones that appear always in the same frequency across the cube. Based on this, the reduction process applies a spatial binning on the data cube, in order to reach a desired signal-to-noise ratio on the sky lines. At the same time it creates a median spectrum of the cube. A cross correlation between the median spectrum and the spatially-binned sky regions allow us to determine the sky dominated regions, which are used to produce a sky cube. Finally, the sky cube is subtracted from the original observed data cube. We refer the reader to TF14 for more details.

\subsection{VV304a residual $V_{\rm los}$ field}
\label{sec:2.3}

In the top panel of Figure~\ref{fig:3} we show the  $V_{\rm los}$ field  of VV304a, derived from the $H_{\alpha}$ data cube. Note that $H_{\alpha}$ emission in this galaxy can be detected up to its optical radius (see Figure B1 of T14). Thus it can be used to map the kinematics  of the disc's outskirts.
The kinematic position angle PA $= 104^{\circ} \pm 2^{\circ}$, inclination angle $i = 39^{\circ} \pm 12^{\circ}$, systemic velocity $V_{\rm sys} = 3811 \pm 3$ km s$^{-1}$, and rotation curve were estimated by fitting an axisymmetric velocity field model to the observed data and minimizing the residual velocity dispersion. This was performed by applying the fitting method developed by \citet[][E08]{2008MNRAS.388..500E}. We refer the reader to E08 for more details about the fitting procedure.  The axisymmetric model for the velocity field is also shown in this panel by the smooth coloured background in regions without observable $H_{\alpha}$.

The observed $V_{\rm los}$ field contains contributions from the three different velocity components: the distributions within the disc plane, radial $V_{\rm R}$ and  rotational, $V_{\phi}$; and the perpendicular velocity  to the disc plane, $V_{\rm Z}$. To remove the modelled unperturbed $V_{\phi}$ velocity field from the observed $V_{\rm los}$ distribution, we subtract the axisymmetric kinematic model from the data. The residual velocity map, $V_{\rm res}$, is shown in the bottom panel of Figure~\ref{fig:3}.  The map shows a strong, global and coherent pattern extending all the way to the disc outskirts, with $V_{\rm res }$ values that can be as large as 50 km s$^{-1}$.

The structure of the residual map is consistent with what is expected for the $V_{\rm Z}$ field of a corrugated disc galaxy. As shown by several previous studies \citep[see e.g.][]{2016MNRAS.456.2779G,2016ApJ...823....4D,2017MNRAS.465.3446G,2018MNRAS.481..286L} vertical velocity flows both in the disc stellar and gas components, caused by a recent interaction with an external perturber, are common and can reach these large amplitudes. These studies have also shown that vertical patterns in the cold star-forming gas and stellar 
components of a disc can remain coincident for more than 1 Gyr \citep{2017MNRAS.465.3446G}. Given the large amplitudes of the perturbations observed in the VV304a $V_{\rm res}$ field ($\sim 50$ km s$^{-1}$)  a significant contribution from off-plane velocity flows, V$_{\rm Z}$, induced by the recent interaction with VV304b, can thus be expected. 
The upper half of the $V_{\rm res}$ map shown in the middle panel of Figure~\ref{fig:3} shows a particularly interesting feature,
which covers approximately $180^{\circ}$ of the galactic disc, extends to the outskirts of the disc, and has an amplitude of $\sim 40$ km s$^{-1}$. Note how the disc's $V_{\rm res}$ sharply transitions from large positive to negative values. Such behaviour 
could indicate a  significant displacement of the disc's $\langle Z \rangle$ with respect to its overall mid-plane. A similar 
displacement has been observed in the outer Milky Way disc. Known as the Monoceros ring, this large and 
complex stellar structure exhibits a north-south asymmetry in $\langle Z \rangle$, with the southern and 
northern parts dominating the regions closer and farther from the Sun, respectively \citep[e.g.][]{2014ApJ...791....9S}.

 However, as previously discussed, due to the non-negligible inclination of the VV304a disc and to the presence of spiral structure and a possible weak bar, the resulting $V_{\rm res}$  could contain  significant contributions from in-plane velocity flows.  \cite{2016MNRAS.461.3835M} studied the effects of bar-spiral coupling on stellar kinematics considering bars and spiral structure, similar to those observed in the Milky Way. They find that the in-plane velocity perturbations associated with these m=2 patterns are $< 20$ km s$^{-1}$ \citep[see also][]{1993ApJ...Canzian}.

It is worth recalling that VV304a does not possess any strong $m=2$ bar perturbation (see Sec.~\ref{sec:2-1}). However, it does show a flocculent-like spiral structure. To quantify the strength of this spiral structure, we generate a 2D model of the galaxy's light distribution. We fit elliptical isophotes to the Gemini/GMOS r'-band image of the galaxy using the IRAF task ELLIPSE with a fixed ellipticity of 0.2. The 2D model is obtained using the isophotal results. Once the model is obtained, we generate a residual overdensity map by assuming a constant mass-to-light ratio across the disc. This is done by computing
\begin{equation*}
    \Delta\rho = \dfrac{\rm image - model}{\rm model}.
\end{equation*}
Note that, in this photometric band, departures from a constant mass-to-light ratio across the disc are expected to be small (see appendix A for more details.)

\begin{figure}
    \centering
    \includegraphics[width=80mm,clip,angle=0]{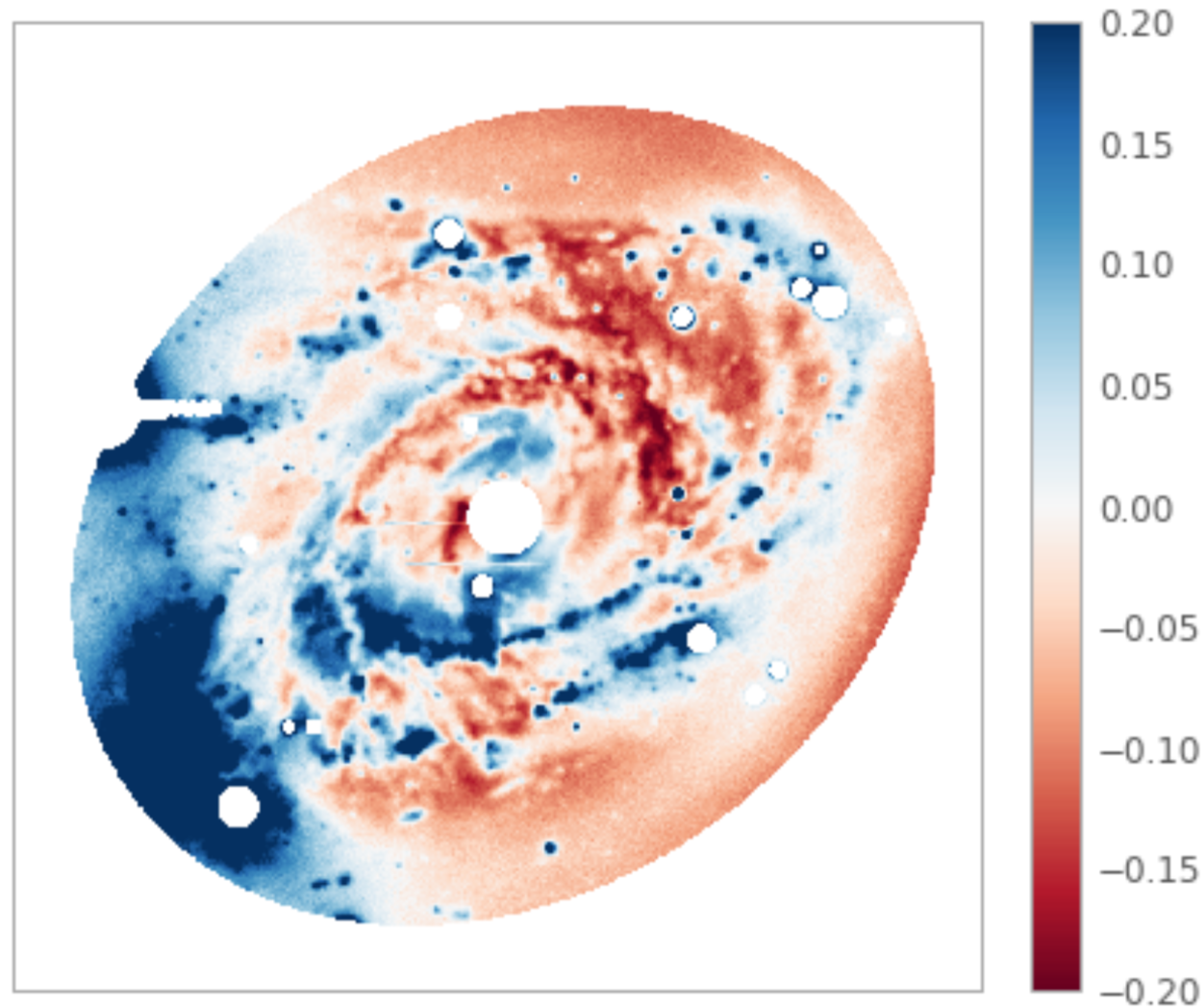}
    \caption{Estimated overdensity map of VV304a. Tho color bar indicates the value of $\Delta\rho =$ (image - model)/model. The light distribution model was obtained by fitting elliptical isophotes to the galaxy's Gemini/GMOS r'-band image. The spiral structure of VV304a shows a density contrast with respect to the background disc of the order $\Delta\rho \lesssim 30$ per cent.}
    \label{fig:overdense}
\end{figure}

The results of this procedure are shown in  Figure~\ref{fig:overdense}, where the color coding indicates the value of $\Delta\rho$. The overdense region in the bottom left area of this figure is likely the result of the recent tidal interaction with VV304b, and can also be appreciated in the top panel of Fig.~\ref{fig:1}. On the other hand, the underdense region in the top right area is the result of significant dust extinction (see also top panel of Fig.~\ref{fig:1}). In general, we find that the spiral structure of VV304a shows a density contrast with respect to the background disc of the order $\Delta\rho \lesssim 30$ per cent, similar to that observed in the MW \citep[see][and references therein]{2016MNRAS.461.3835M}. In the following Section we will use this information to further characterize the contribution from in-plane velocity flows that could arise from bars or bi-symmetric spiral structure to the $V_{\rm res}$ field of a low-inclination disc.

\begin{figure*}
    \centering
    \includegraphics[width=180mm,clip,angle=0]{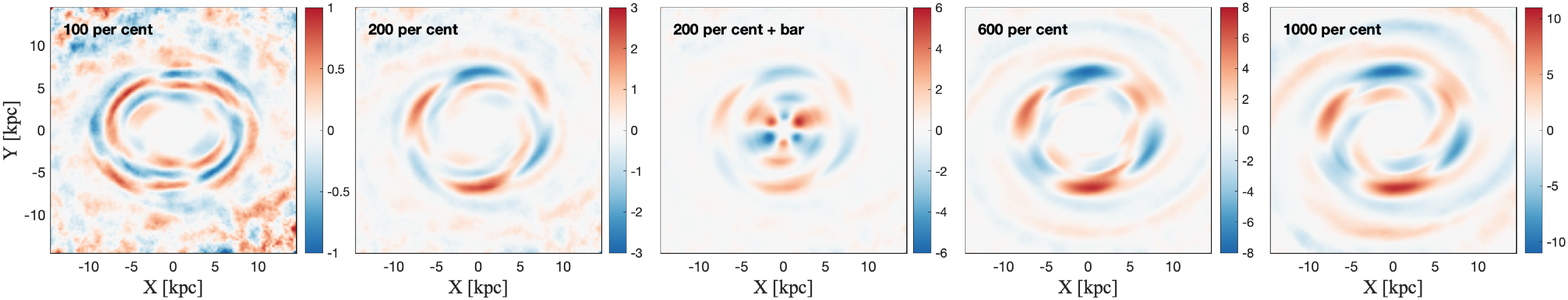}
    \includegraphics[width=180mm,clip,angle=0]{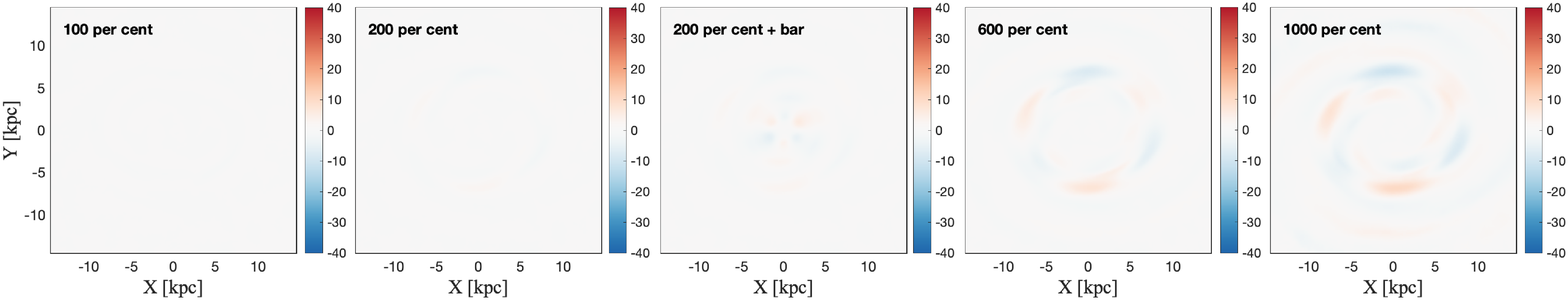}
\caption{$V_{\rm los}$ maps from our suite of test particle simulations, obtained after  rotating the models to emulate the orientation of VV304a, ($i =35^{\circ}$,~PA = $105^{\circ}$). The top left corner of each panel indicates the value of the density contrast at $R_0$ of the spiral structure, $\Delta\rho$, with respect to the background disc surface density. The third panel also includes the contribution from a galactic bar with similar parameters to those of the MW bar. The color bar in the top panels is scaled to the corresponding maximum $V_{\rm los}$ value. For comparison, the color bars in the bottom panels have been scaled as in the bottom panel of Fig.~\ref{fig:3}. Note that even our our strongest spiral model, $\Delta\rho = 1000$, cannot generate in-plane velocity perturbations as large as those seen in VV304a.}
    \label{fig:analit}
\end{figure*}

\section{Analytic models of Perturbed velocity fields}

It is well known that radial and tangential flows  can arise from the dynamical influence of a bar or spiral structure on a disc \citep{2012MNRAS.425.2335S,2016MNRAS.460L..94G,2016MNRAS.461.3835M}, and that the amplitudes of such flows vary with the strength of these $m=2$ modes. To  characterize the contribution from in-plane velocity flows driven by the VV304a non-axisymmetric structure, we have run and analyzed a suite of test particle simulations using a MW-like galactic potential. The model parameters were chosen to reasonably reproduce the properties of VV304a listed in Sec.~\ref{sec:2-1}.

The analytic galactic model considers a Hernquist profile for the bulge component,
\begin{equation}
\Phi_{\rm bulge}=-\frac{GM_{\rm bulge}}{\sqrt{x^2+y^2+z^2}+\epsilon_{\rm bulge}},
\end{equation}
with $M_{\rm bulge}=0.47\times10^{10} {\rm M}_{\odot}$ y $\epsilon_{\rm bulge}=0.83 {\rm Kpc}$. To approximate a double exponential disc profile, we follow the procedure described by \citet{2015MNRAS.448.2934S}. The idea behind this method is to approximate an exponential profile by the superposition of three different Miyamoto \& Nagai (MN) profiles \citep{1975PASJ...27..533M}. \citet{2015MNRAS.448.2934S} provides a user-friendly online web-form that computes the best-fitting parameters for an exponential disc\footnote{\url{http://astronomy.swin.edu.au/~cflynn/expmaker.php}},
\begin{equation}
\rho(R, z) = \rho_{0} \exp(-R/\epsilon^{\rm s}_{\rm disc})\exp(-|z|/\epsilon^{\rm h}_{\rm disc})\ ,
\end{equation}
with $\rho(R, z)$ the axysimmetric density, $R=\sqrt{r^2-z^2}$ the projected galactocentric distance, $\rho_0$ the central density. Our resulting disc model has a total mass $M_{\rm disc}=5.9\times10^{10} ~{\rm M}_{\odot}$, scalelength $\epsilon^{\rm s}_{\rm disc} = 3.1$ kpc and scaleheight $\epsilon^{\rm h}_{\rm disc} = 0.3$ kpc. For the dark matter halo we consider a NFW profile
\begin{equation}
\Phi_{NFW}=-\dfrac{A}{r}\ln\left[1+\dfrac{r}{r_s}\right],
\end{equation}
where
\begin{equation}
\nonumber
A=\dfrac{GM_{200}}{\ln(1+c_{{\rm NFW}})-\frac{c_{{\rm NFW}}}{1+c_{{\rm NFW}}}}, 
\end{equation}
with  $M_{200}=1.45 \times 10^{12} {\rm M}_\odot$, $r_{200}=234.5 ~{\rm kpc}$ y $r_s=14.6 ~{\rm kpc}$. Note that these parameter choices yield a total dynamical mass within 18 kpc of $\approx 2.55 \times 10^{11}$ M$_{\odot}$ and peak circular velocity of 250 km/s, in good agreement with VV304a.

Following \citet[][hereafter M16]{2016MNRAS.461.3835M} we include a 3D bar potential, described by
\begin{equation}
\Phi_{\rm bar}(R,\phi,z,t)=\alpha\frac{v_0^2}{3}\left(\frac{R_0}{R_b}\right)^3U(r)\frac{R^2}{r^2}\cos(\gamma_b),
\end{equation}
where $R_b$ is the length of the bar, $v_0$ is the circular velocity at $R_0$, $\gamma_b(\phi,t)\equiv 2\left(\phi-\phi_b-\Omega_b\;t\right)$, and
\begin{equation}
U(r)\equiv\left\{ \begin{array}{lcl}
-(r/R_b)^{-3} & {\rm for } & r\ge R_b,\\
(r/R_b)^{3}-2 & {\rm for } & r< R_b.
\end{array}
\right.
\end{equation}
The amplitude $\alpha$ is the ratio between the bar's and axisymmetric contribution to the radial force along the bar's long axis at $(R, z) = (R_0, 0)$. Even though, as previously shown, VV304a does not present a well defined bar perturbation, we included in our suite a model with a bar with similar characteristic to that of the MW. This is $\alpha= 0.01$ , $\Omega_b = -52.2~{\rm km ~s^{-1}}~{\rm kpc}^{-1}$, and $R_b = 3.5~{\rm kpc}$. This last model also includes a spiral perturbation with a 200 per cent density contrast. 

Finally, as in M16, we introduce a 3D spiral perturbation consisting in a two--armed model as proposed by Cox \& G\'omez (2002),

\begin{equation}
\Phi_{\rm spiral}(R,\phi,z,t)=-\frac{A}{R_s KD}\cdot e^{-\frac{\left(R-R_s\right)}{R_{\rm sd}}}\cos(\gamma_s)\left[{\rm sech}\left(\frac{Kz}{\beta}\right)\right]^\beta,
\end{equation}
where

\begin{equation}
\begin{array}{l}
K(R)=\dfrac{2}{R\sin(p)},\\ \\
\beta(R)=K(R)h_s\left[1+0.4K(R)h_s\right],\\ \\
D(R)=\dfrac{1+K(R)h_s+0.3\left[K(R)h_s\right]^2}{1+0.3K(R)h_s},\\ \\
\gamma_s(R,\phi,t)=2\left[\phi-\phi_s-\Omega_s t-\frac{\ln(R/R_s)}{\tan p}\right].
\end{array}
\end{equation}
Here, $p$ is the pitch angle, $A$ the amplitude of the spiral potential, $h_s$ controls the scale--height of the spiral, and $R_s$ is the reference radius for the angle of the spirals. We choose $R_s = 1~{\rm kpc}$, $R_{sd} = 3.1~{\rm kpc}$, $\Omega_s = -18.9~{\rm km s}^{-1} ~{\rm kpc}^{-1}$ and $p = 9.9^\circ$. For our study, the most relevant parameter is $A$, which sets the strength of the perturbation. When modelling the MW M16 considered two values of $A$. The first, $A = 341.8$ km$^2$ s$^{-2}$ and  referred to as `reference spirals', corresponds to a 30 per cent density contrast of the spiral arms with respect to the background disc surface density at  $R = R_0 = 8$ kpc. A second model, referenced to as `strong spirals', considers  $A = 683.7$ km$^2$ s$^{-2}$; i.e. a 60 per cent density contrast. These values of $A$ result on a maximum radial force exerted by the spiral arms  of 0.5 per cent and 1 per cent of the force due to the axisymmetric background at $R_0$, and introduced velocity perturbation in $V_R$ and $V_{\phi} < 15$ km/s. As previously discussed in Section~\ref{sec:2.3}, the strength of VV304a spiral structure  lies in between the M16 reference and strong models. Nonetheless, in this work we are interested in exploring how strong a spiral perturbation should be in order to induce perturbation in the $V_{\rm los}$ field as large as those observed in VV304a (i.e. $\sim 50$ km/s).  For this purpose, we also explore significantly larger values of $A$, namely a 100, 200, 600 and 1000 per cent  density contrast, $\Delta\rho$, with respect to the background disc surface density at $R_0$. The later model translate into a maximum radial force exerted by the spiral arms  of $\sim 17\%$ of the force due to the axisymmetric background at $R_0$.

For each model a set of $\sim5 \times 10^6$ test particles,  sampling of the phase-space distribution of the unperturbed stellar disc, were initialized in equilibrium using the publicly available code \texttt{MAGI}\footnote{\url{ https://bitbucket.org/ymiki/magi}} \citep{2018MNRAS.475.2269M}. The initial conditions were integrated using another publicly available code, the \texttt{LP-VIcode}\footnote{\url{http://lp-vicode.fcaglp.unlp.edu.ar/}} \citep{2014A&C.....5...19C} to make them evolve for 9.6 Gyr to allow the system to relax. This integration time was chosen so that any spurious substructure on the disc density field, arising due to the unrelaxed set of initial conditions, are well mixed within the inner 20 kpc. 
After that period of time, the non-axisymmetric perturbations to the potential were slowly introduced by linearly increasing in time their amplitude to the desired value. Following \citet{2016MNRAS.461.3835M}, this was done within a period of 3 Gyr, which represents $\approx 15$ full rotations of non-axisymmetric perturbations. Finally, each model was allowed to evolve under the influence of the perturbed potential for an additional period of 3 Gyr. 

In Figure ~\ref{fig:analit} we show the resulting $V_{\rm los}$ map for each model, after rotating them into an inclination of $35^{\circ}$ and a position angle of $105^{\circ}$ to emulate VV304a orientation. To compute the $V_{\rm los}$ we subtracted from each particle's $V_{\phi}$ the circular velocity at the corresponding galactocentric radii. To highlight the effect of the non-axysimmetric perturbation, the colorbars in the top panels  have been scaled based the maximum $V_{\rm los}$ value obtained on each map. The left most panel shows the results obtained for the $\Delta\rho = 100\%$ model, similar to the strong spiral model of M16 ($\Delta\rho = 60\%$). As expected, our results are in good agreement with those of M16 (see their Fig. 5). After inclining the disc to $i=35^{\circ}$, the resulting perturbation in the $V_{\rm los}$ field is of the order of  only $1$ km/s. As we increase the strength of the spiral perturbation, the magnitude of the in-plane flows significantly increases, reaching values of 10 km/s for our most extreme model, i.e. $\Delta\rho = 1000\%$. The second and third panels show a comparison between models without and with bars, respectively. Both models include a spiral perturbation with $\Delta\rho = 200\%$. Note that the bar enhances the magnitude of the in-plane flows in the very inner disc regions, doubling its amplitude with respect to the barless model. Yet, even with the bar included, the $V_{\rm los}$ perturbations reach values $< 6$ km/s. In the bottom panels of Figure~\ref{fig:analit}, for comparison, we show the same models but now the color bars have been set as in the bottom panel of Figure \ref{fig:3}. This figure clearly shows that, for a galaxy with  $i = 35^{\circ}$, neither a bar as strong as the one of the MW nor spiral structure with a $\Delta\rho$ as large as 1000 per cent can secularly introduce perturbations in the $V_{\rm los}$ field with amplitudes as large as those observed in VV304a. We recall that VV304a does not show a bar and that the $\Delta\rho$ of its spiral structure is $< 100$ per cent of the background density of its disc.

\begin{figure}
    \centering
    \includegraphics[width=50mm,clip,angle=90]{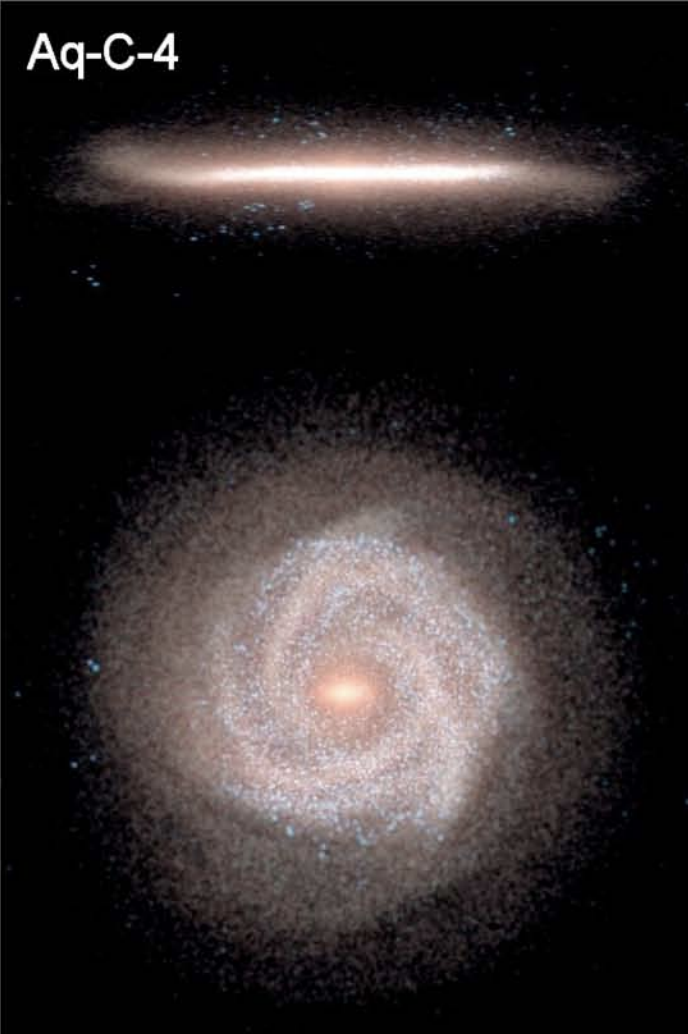}
    \caption{Present-day stellar distribution of the Aq-C4 disc. The images were constructed by mapping the K-, B- and U-band luminosities to the red, green and blue channels of a full colour composite image. Younger (older) star particles are therefore represented by bluer (redder) colours. The left and right panels show the edge-on and face-on views, respectively. Note the simulated disc shows a weak bar and multi-arm spiral structure, similar to what is observed in VV304a disc. Figure from \citet{2014MNRAS.437.1750M}.}
    \label{fig:c4}
\end{figure}

In this Section we have analyzed the secular response of a disc to a bar and spiral perturbation using test particle simulations of an isolated galaxy. In what follows we use highly resolved fully cosmological self-consistent hydrodynamical simulations to characterize the impact that in-plane velocity flows, associated with both strong and mild $m=2$ modes, can have on the $V_{\rm res}$ field of 
a galactic disc that has recently interacted with an external perturber.

\section{Self consistent models of perturbed velocity fields}
\label{sec:numsim}

In the previous section we used a suite of test particles simulations to show that the VV304a spiral structure  can not secularly drive in-plane flows sufficiently large to account for the velocity perturbation observed in its  $V_{\rm res}$ field. In this section we further study  the impact that in-plane flows may have on the $V_{\rm res}$ field of this low-inclination interacting galaxy by analyzing fully self-consistent cosmological simulation of VV304a like models. We have focused on models that have recently interacted with an external companions and that present {\it i)}  weak and {\it ii)} very strong $m=2$ modes.  

\begin{figure}
    \centering
    \includegraphics[width=85mm,clip,angle=0]{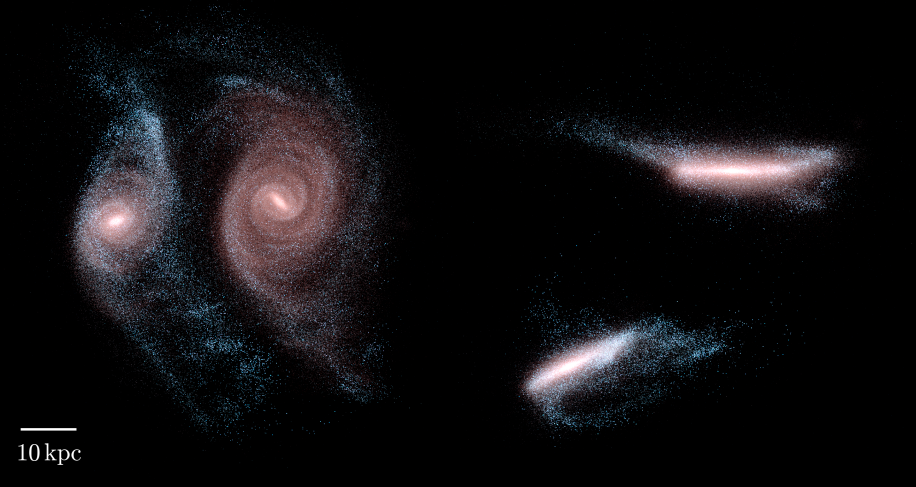}
    \caption{The moment of closest approach between the simulated Auriga galaxy, Au25, and its massive companion. The images were created
    as in Fig.\ref{fig:c4}.  The face-on (left) and edge-on (right) views are oriented with respect to the Au25 disc. The disc shows a strong bar and a grand design spiral structure.}
    \label{fig:5}
\end{figure}

The main advantage of using fully cosmological simulations, as opposed to 
tailored simulations of interacting galaxies set in isolation, is that the former 
self-consistently include mechanisms that can significantly perturb an otherwise 
initially relaxed disc, such as previous close encounters with satellites,
distant flybys of massive objects, and accretion of misaligned cold gas from halo 
infall or from mergers. Thus a realistic representation of possible sources of confusion
is included in our study.




\subsection{The Auriga models}

The Auriga project \citep{2014MNRAS.437.1750M,2017MNRAS.467..179G,2019MNRAS.490.4786G} consists of $\sim 40$ 
high-resolution cosmological zoom-in simulations of the formation  of 
late-type  galaxies  within  Milky  Way-sized  halos,  performed  with  the 
N-body,  magneto-hydrodynamics  code {\sc AREPO} \citep{2010MNRAS.401..791S}. 
The code includes modelling of a wide range of critical physical processes that govern galaxy formation \citep{2014MNRAS.437.1750M}
such as gas cooling/heating, star formation, 
mass return and metal enrichment from stellar evolution, the growth of a supermassive black hole, and feedback from stellar 
sources and from black hole accretion. The 
halos were selected from a lower resolution dark matter only simulation from
the Eagle Project, a periodic cube of side length 100 comoving Mpc 
\citep{2015MNRAS.446..521S}.  Each halo was chosen to have,  at $z= 0$,  a virial mass in 
the range of $0.5 - 2 \times 10^{12}$ M$_{\odot}$ and to be more distant than nine times the virial radius from any 
other halo of mass more than 3\% of its own mass. The dark matter particle and 
gas cell mass for the  simulation considered here are $\sim 3 \times 10^{5}$ M$_{\odot}$ and $\sim 5 \times 10^4$ 
M$_{\odot}$, respectively.  The gravitational softening length of the stars and dark matter grows with scale factor up to a maximum of 
369 pc, after which it is kept constant in physical units. The softening length of gas cells scales with the mean radius of the cell, but is never allowed to drop below the stellar softening length. A study across three resolution levels shows that many galaxy 
properties, such as surface density profiles, orbital circularity distributions, star formation histories and disc vertical 
structures are already well-converged at the resolution level considered in this work.

Previous studies based on the same simulations have shown that vertical patterns in the cold star-forming gas and stellar 
components of a disc can remain coincident for more than 1 Gyr \citep{2017MNRAS.465.3446G}. Thus, we focus on the stellar components which have 
much better spatial resolution compared to the cold gas discs. Nevertheless, it is worth noticing that the vertical pattern in the two models considered here can also be recovered from the cold star-forming gas distribution.

\begin{figure}
    \centering
    \includegraphics[width=60mm,clip]{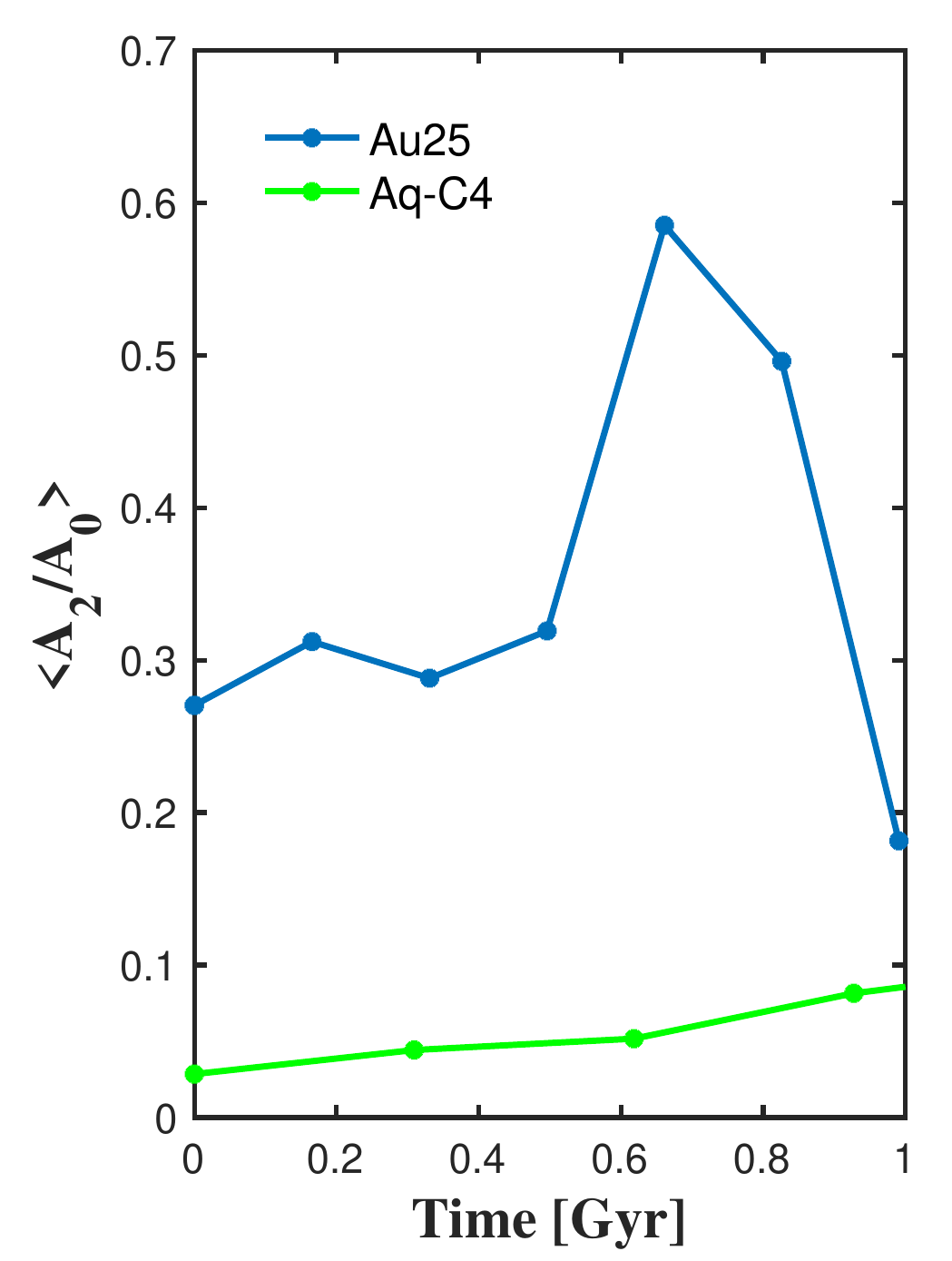}
    \caption{The time evolution of the amplitudes of the $m=2$ Fourier mode of the simulated stellar discs analyzed in this work. The green and blue lines
    show  results for the Aq-C4 and Au25 discs, respectively.}
    \label{fig:m=2}
\end{figure}

\begin{figure}
    \centering\textit{
    \includegraphics[width=90mm,clip]{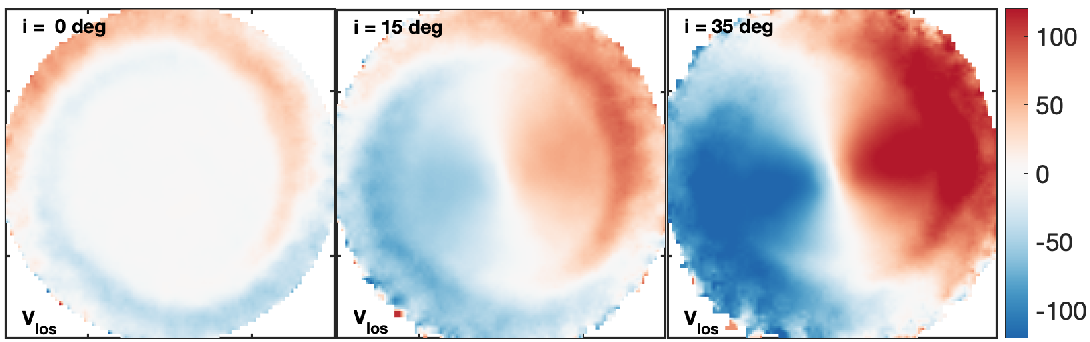}
    \includegraphics[width=90mm,clip]{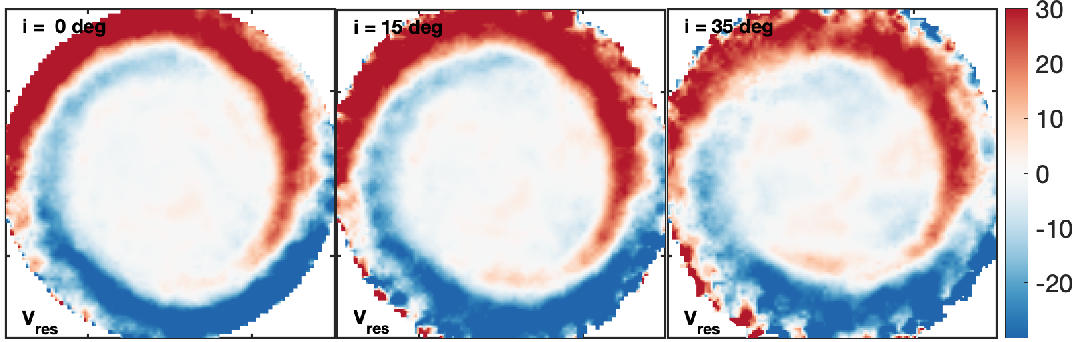}
    \includegraphics[width=90mm,clip]{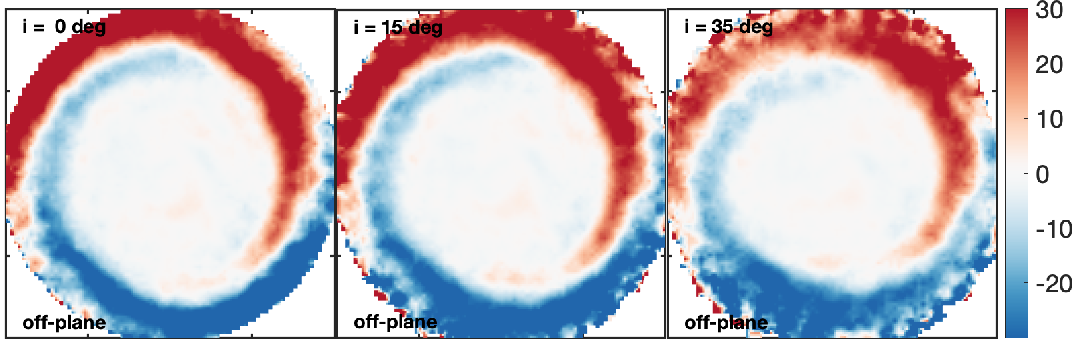}
    \includegraphics[width=90mm,clip]{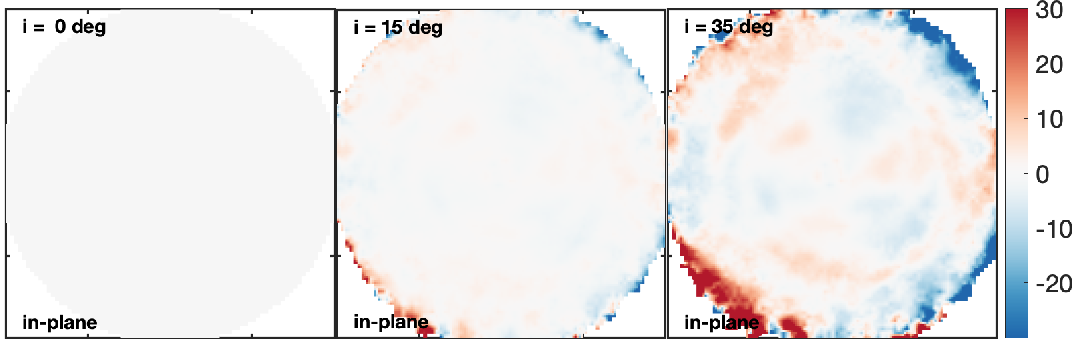}}
    \caption{First row: Aq-C4 line-of-sight velocity field, $V_{\rm los}$, obtained for three different inclination angles. 
    From left to right we show the  maps after tilting the disc by $i=0^{\circ},15^{\circ}$ and $35^{\circ}$, respectively. 
    The second row shows residual velocity maps, $V_{\rm res}$, obtained after subtracting from every stellar particle the mean rotation velocity at the corresponding radial position. The third and fourth rows show the contributions from vertical and in-plane motions to the $V_{\rm res}$ fields,
    respectively. Note that even at $i=35^{\circ}$   $V_{\rm res}$ is dominated by vertical motions, $V_{\rm Z}$.}
    \label{fig:tt}
\end{figure}

\subsubsection{Weak $m=2$ model}
\label{sec:weak}

We first analyze the Aq-C4 model, first introduced in \citet{2014MNRAS.437.1750M}. Aq-C4 has the property of being 
one the most similar models to the Milky Way within the sample. It has a flat rotation curve
that peaks at 250 km s$^{-1}$ at $\sim$ 4 kpc. Its disc scale length and bulge effective radius
are approximately 3.1 and 0.8 kpc, respectively. At the present-day it has $M_{\rm tot} \sim 1.46 \times
10^{12}$ M$_{\odot}$,  baryonic disc + bulge mass  $M_{*} \sim 5.3 \times 10^{10}$ M$_{\odot}$ and a 
disc optical radius  of $\sim 20$ kpc. Here we define the optical radius as the radius where 
$\mu_{\rm B} = 25$ mag arcsec$^{-2}$ \citep[cf.][]{2010MNRAS.406..576P}. Note that these properties make Aq-C4 a very reasonable model of VV304a (see Sec.~\ref{sec:2-1}). Aq-C4 shows a quiet formation history since $z\approx 1$, without any close interaction 
with satellites more massive than $M_{\rm sat} / M_{\rm host} = 1/40$.  The most significant perturber is a low-velocity
fly-by with a pericentre passage of 80 kpc at a lookback time of $t_{\rm look} \approx 2.7$ Gyr. The satellite
has a total mass of $\sim 4 \times 10^{10}$ M$_{\odot}$ ($M_{\rm sat} / M_{\rm host} \approx 1/50$). As shown 
in \citet[][hereafter G16]{2016MNRAS.456.2779G}  the satellite is not massive enough to directly perturb the galactic
disc but the density field of the host dark matter halo responds to this interaction resulting in a strong amplification 
of the perturbative effects. The resulting dark matter  wake is efficiently transmitted to the inner parts of the host 
perturbing the embedded disc. This perturbation causes the onset and development of a vertical pattern, similar to 
the Monoceros ring. Figure~\ref{fig:c4} shows the present-day stellar distribution. The disc has a weak bar and weak multi-arm  
spiral structure, similar to that observed in the VV304a galactic disc. The  green line in Figure~\ref{fig:m=2}, 
shows the time evolution of the amplitude of the $m=2$ mode in Aq-C4's disc, computed within the inner 20 kpc following the approach of 
 \citet{2016MNRAS.459..199G}.

\subsubsection{Strong $m=2$ model}
\label{sec:strong}

Among the Auriga models, Au25 is the one that most closely resembles the VV304 system in terms of its global properties. Furthermore, as discussed below, it has the strongest $m=2$ close to $z=0$. The host galaxy has, at the present-day, a 
total mass of $M_{\rm tot} \sim 1.25 \times 10^{12}$ M$_{\odot}$, a baryonic disc + bulge mass of $M_{*} \sim 3.4 \times 10^{10}$ M$_{\odot}$, a peak circular 
velocity of $\sim 180$ km s$^{-1}$, and a disc optical radius that azimuthally varies between 21 and 30 kpc. Even though Au25 is slightly less massive than VV304a, it had a close encounter with a massive 
satellite 0.9 Gyr ago whose total mass at infall was $M_{\rm sat} \sim 3 \times 10^{11}$ M$_{\odot}$ and whose orbit reached the host's inner 40 kpc. This configuration is reminiscent of the  recent interaction in VV304a and b. Figure~\ref{fig:5} highlights the moment of closest approach. During the interaction, the host galaxy is strongly tidally perturbed and develops a significant misalignment of its
outer dark matter halo regions. The interaction also induces the formation of two tidal arms and triggers strong vertical patterns whose signatures can still be 
detected clearly at the present-day \citep{2017MNRAS.465.3446G}.  Note the 1:4 host to satellite mass ratio, which can be classified as a massive interaction.
In Figure~\ref{fig:m=2} we show the time evolution of its $m=2$ mode, measured as in Sec.~\ref{sec:weak}. 
The interaction with its massive companion triggers the 
development a strong $m=2$ signal associated with both a galactic bar and grand design spiral structure (see Fig.~\ref{fig:5}). \citet{2016MNRAS.459..199G} 
studied the time evolution of the $m=2$ modes on a large sample of the Auriga models. 
Figure 4 of that work shows that Au25 has the strongest $m=2$ mode during the last 3 Gyr of evolution. 

\subsection{Simulated residual $V_{\rm los}$ fields}
\label{sec:meth}

\begin{figure}
    \centering \textit{
    \includegraphics[width=90mm,clip]{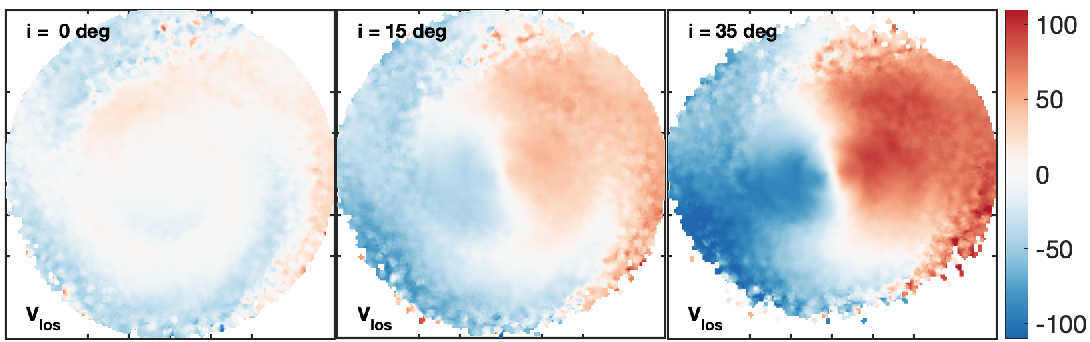}
    \includegraphics[width=90mm,clip]{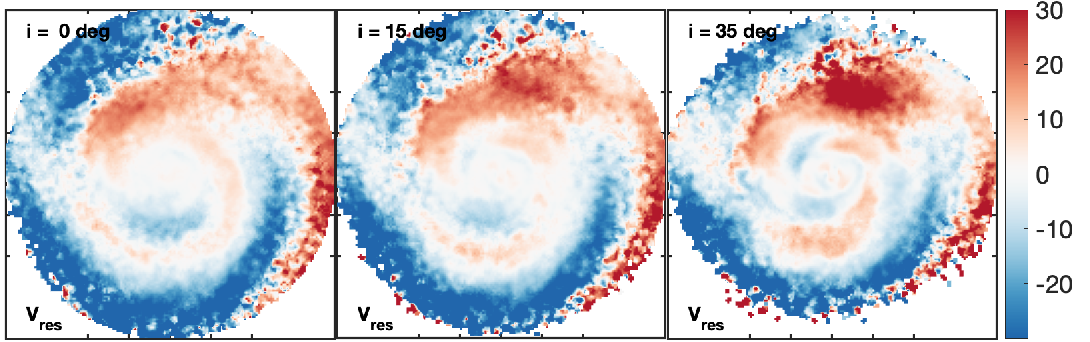}
    \includegraphics[width=90mm,clip]{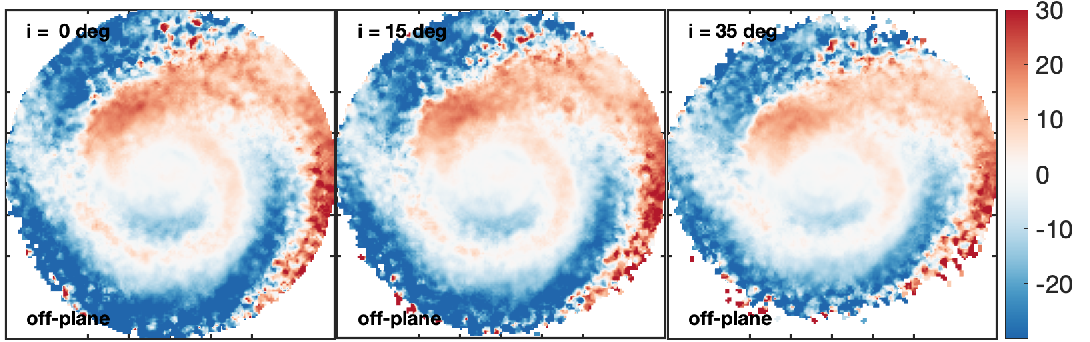}
    \includegraphics[width=90mm,clip]{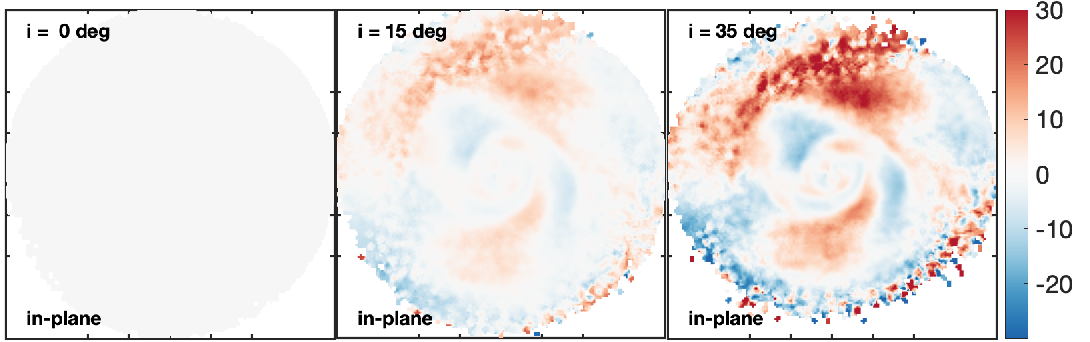}
    }
    \caption{As in Figure~\ref{fig:tt}, for Au25.}
    \label{fig:tt2}
\end{figure}

To investigate whether our simulated discs develop similar velocity patterns to those observed in VV304a, we first 
rotate the main host discs to an inclination of $i=35^{\circ}$ and $PA = 100^{\circ}$, similar to the values estimated for VV304a. We then compute 
maps of mean $V_{\rm los}$ across the disc plane for all snapshots after the interactions. The top rightmost panels of Figures~\ref{fig:tt}
and \ref{fig:tt2} show an example of the $V_{\rm los}$ map obtained from the Aq-C4 and Au25 discs, respectively. The snapshot 
corresponds to $z=0$ for Aq-C4, 2.4 Gyr after 
the flyby closest approach, and $\sim 0.75$ Gyr after the strong interaction suffered by Au25, just one snapshot before $z=0$. 
Although some interesting departures from perfectly rotating discs can be observed, in both cases the resulting maps are clearly dominated by the $V_{\phi}$ velocity component.

To obtain the $V_{\rm res}$ velocity field  for the simulated stellar discs, we 
subtract from every stellar particle the mean rotation velocity at the corresponding radial position. 
The resulting maps are shown on the second row rightmost panels of Figs.~\ref{fig:tt}
and \ref{fig:tt2}. As in the VV304a case, the $V_{\rm res}$ maps show strong and 
coherent velocity patterns that can reach amplitudes as large as 50 km s$^{-1}$. 
Note that both maps show similarly strong flows, despite the differing 
strength of their in-plane disc $m=2$ modes. The morphology of the coherent
flows show very similar features to those 
 in the observational data, with sharp transitions between positive and negative $V_{\rm res}$ velocities. Such transitions are associated with extrema 
 in the displacement of the mid-plane of the discs if the $V_{\rm res}$ fields are indeed dominated by off-plane motions, $V_{\rm Z}$.
To explore this, we show on the second row  of Figs~\ref{fig:tt} and \ref{fig:tt2} the $V_{\rm res}$ maps obtained after
inclining the disc by only $15^{\circ}$ (second panel) and $0^{\circ}$ (first panel). 
Note that in a perfectly  face-on configuration, i.e. $i = 0^{\circ}$, the resulting $V_{\rm los}$
and $V_{\rm res}$ maps are equivalent to the disc's mean vertical velocity distribution, $V_{\rm Z}$. Thus, coherent and extended departures from
0 km s$^{-1}$ on these maps are a direct indication of a corrugation in the simulated host discs. 

It is clear  in both cases that 
the patterns seen in  the $V_{\rm res}$ field are preserved as we tilt the discs towards lower inclinations. 
In the case of Aq-C4, these are not only preserved but the amplitude of the patterns increases for lower inclination angles. This indicates
a negligible contribution from in-plane flows to the $V_{\rm res}$ field, even at $i=35^{\circ}$. This is not surprising 
since, as in the case of VV304a,  the simulated disc shows a very mild $m=2$ pattern. 
A similar result is found for Au25 (see Fig.~\ref{fig:tt2}). The global morphology of the 
$V_{\rm res}$ field is preserved as we decrease the inclination. 
The $i=0^{\circ}$ map reveals very strong vertical flows, with amplitudes larger than 40 km s$^{-1}$,
that are the result of a disc corrugation. Note the trailing nature of this corrugation pattern. Contrary to the Aq-C4 case, where leading 
patterns develop over time due to the inner disc torque \citep[][]{2006MNRAS.370....2S, 2016MNRAS.456.2779G}, 
the vertical structure of the Au25  disc traces the trailing structure of the tidal arms created during the recent interaction. 

It is clear that vertical flows dominate the $i=35^{\circ}$ $V_{\rm res}$ field in Au25 disc. 
However the amplitudes of some of these features decrease for
lower inclination, indicating a non-negligible contribution from in-plane flows. 
To further characterize the contribution from the different velocity components,  we show in  Figs~\ref{fig:tt} 
and \ref{fig:tt2}  velocity maps in which only vertical (third row) and in-plane (fourth row) motions are included.  
For Aq-C4 (weak $m=2$ mode) it is clear that the contribution from in-plane flows is negligible, even at $i=35^{\circ}$. In the Au25 case 
(strong $m=2$), even though significant in-plane flows can be found at certain locations, the off-plane flows dominate the total residual velocity patterns over most of the disc, even at $i=35^{\circ}$. This shows that the velocity perturbations seen in the  
inclined simulated disc are indeed primarily associated  with a corrugation pattern.

Clearly, features  in the $V_{\rm res}$ map of a low-inclination disc that has recently interacted with 
a companion can be dominated by flows in the direction perpendicular to the disc plane. We recall that this is true even for Au25 which
shows the strongest in-plane $m=2$ pattern in the Auriga sample close to $z=0$. The disc of VV304a  shows neither a strong bar nor a grand design spiral
and thus more closely resembles the structure seen in Aq-C4. This suggests that the global and coherent velocity patterns seen in VV304a could
indeed be associated with vertical flows. 

 \begin{figure*}
    \centering
    \includegraphics[width=180mm,clip]{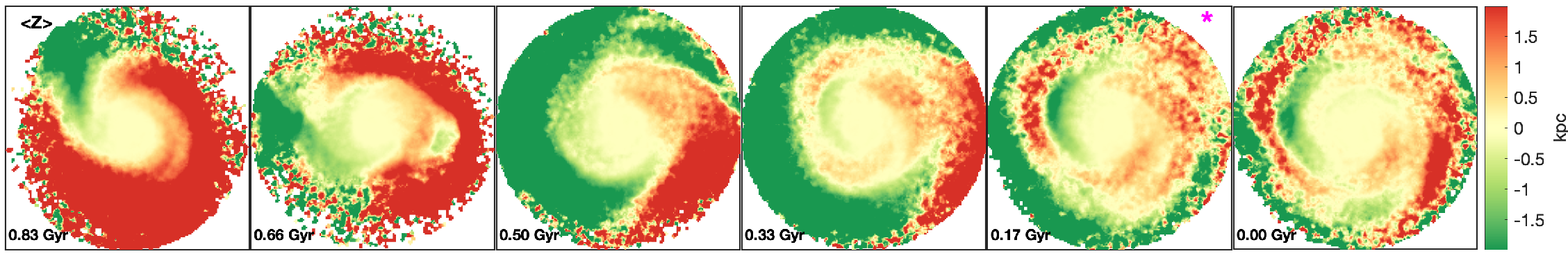}
    \includegraphics[width=180mm,clip]{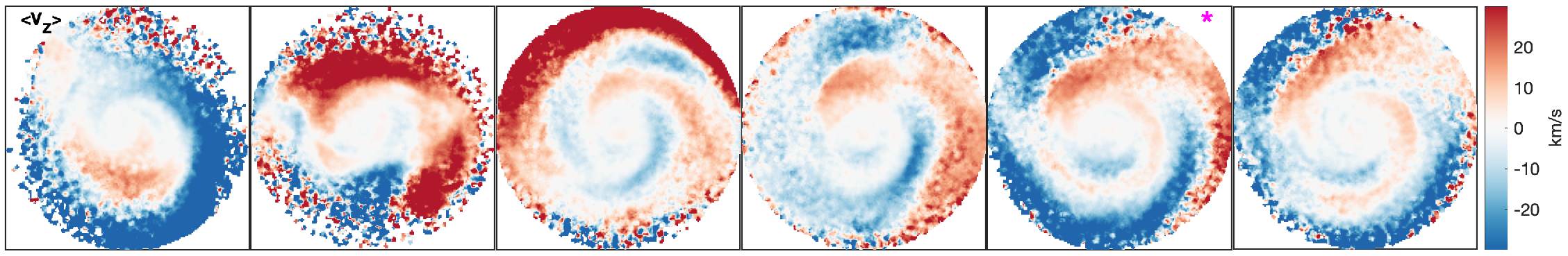}
    \caption{Time evolution of the vertical structure of the disc in Au25 over a period of $\sim 1$ Gyr. The top panels show maps
of the mass-weighted $\langle {\rm Z} \rangle$. The bottom panels show maps of the mass-weighted  $\langle V_{\rm Z} \rangle$. The 
onset-time of the vertical perturbation  is $1 \lessapprox t^{\rm onset} \lessapprox 0.8$ Gyr, coinciding with the pericenter passage of the most massive satellite. The perturbation starts as a $m=1$ mode that winds-up in time. The snapshot used for comparison with the VV304a data is highlighted with the magenta stars(second from the right.
In this projection the galactic disc rotates counterclockwise.}
    \label{fig:mvz}
\end{figure*}

In Figure~\ref{fig:mvz} we show the time evolution of the mean height, $\langle Z \rangle$ (top), and mean 
vertical velocity, $\langle V_{Z} \rangle$ (bottom) of the stellar disc in Au25. A similar figure for Aq-C4 can be found in \citet[][fig. 3]{2016MNRAS.456.2779G}.
This period of time covers the satellite's closest approach and the snapshot used for  data to model comparison, indicated with a magenta star.
The link between the onset of the disc vertical perturbation and closet approach of the most massive companion, at $t \sim 1$ Gyr, 
is evident.  Notice how the vertical perturbation starts as an $m=1$ pattern that winds-up as time goes by. The 
 counterparts of the extended and coherent structures 
 seen in the $\langle V_{Z} \rangle$ and inclined $V_{\rm res}$ fields  at $t = 0.17$ and $0$ Gyr are clearly associated 
 with a  vertical displacement of the disc with respect to the overall  mid-plane. These structures resemble one of the largest vertical 
 asymmetries of our own Galactic disc: the Monoceros ring. A similar feature is seen in the $V_{\rm res}$ field of 
 VV304a (bottom panel of Fig.~\ref{fig:3}), which strongly suggests the presence of a Monoceros ring-like feature in this external late-type 
 galaxy. 
 
The origin of the Monoceros ring has been a puzzle for
many years and it has been debated whether its constituent stars are the debris of a disrupted satellite galaxy or whether the structure
is the projection of a corrugation pattern in our own disc \citep[see][and references therein]{2018MNRAS.481..286L,2018MNRAS.474.4584G}. 

\section{Conclusions}

 In this work we have analyzed the 2D $V_{\rm los}$ field of the late-type galaxy VV304a, obtained through 
$H_{\alpha}$ Fabry-Perot observations. In many aspects, VV304a can be considered a Milky Way analogue which
is currently interacting with a massive companion, VV304b. After subtracting an axisymmetric kinematic model from 
the observed $V_{\rm los}$ field, we find that the residuals show strong, global and coherent motions 
which are consistent with a corrugation pattern. The contribution from in-plane flows to the $V_{\rm res}$ field cannot be subtracted with the available data. However, non-axisymmetric features such as spiral arms and/or a bar are much too weak in VV304a to generate line-of-sight perturbations as large as those measured. We have demonstrated this by analysing multi-band images of VV304a, and comparing to  a suite of test-particle simulations of non-axisymmetric galaxies with bars and m=2 spiral structure but no extraplanar perturbations. Even spiral spiral arms an order of magnitude stronger than those observed in VV304a do not induce velocity flows strong enough to explain the perturbations seen in its $V_{\rm res}$ field. Based on these results, we analysed fully cosmological
models with strong and weak $m=2$ patterns 
to explore whether the $V_{\rm res}$ field of a galaxy that recently interacted with an external perturber 
could still be dominated by signatures of a corrugation pattern, even for a disc as inclined as VV304a. Our results show that this can indeed be the case. 

The VV304a $V_{\rm res}$ map hints at a structure that resembles the Monoceros ring,  
an extended low latitude stellar overdensity in our own Galaxy which may therefore be associated to a global disc corrugation pattern. Several recent studies have linked this structure to interactions with 
the Sgr dwarf galaxy \citep{2011Natur.477..301P,2013MNRAS.429..159G,2018MNRAS.481..286L}, even though its origin is still being debated. Note that 
although  Sgr was likely less massive at infall than VV304b, several different arguments suggest a total infall mass of the order of $\gtrsim 10^{11}~M_{\odot}$ \citep[e.g.][]{2011Natur.477..301P}; this could thus be a 1:10 interaction. As shown in \citet{2017MNRAS.465.3446G}, strong global corrugation patterns can be induced during such interactions.

Our results demonstrate that it is possible to address important questions regarding the nature and origin of vertical perturbations by measuring the velocity distribution of nearby nearly face-on galaxies. What is the prevalence of vertical corrugation patterns in the Local Universe? What are the main physical mechanisms behind the formation of such patterns? And what can they tell us about the recent interaction history of their host galaxies? The tidal origin of  VV304a's perturbation is very clear: a close 
 interaction with VV304b. However, the presence of a corrugation pattern could also be used to constrain the
asymmetries of the host dark matter halo and to study unseen structure in the outskirts of
galaxies \citep{2000ApJ...534..598V,2016MNRAS.456.2779G}.  Our study opens a new window to investigating whether the corrugation pattern observed in the Milky Way disc is a rare or common feature of late-type galaxies. 

\section*{Acknowledgements}

FAG acknowledges financial support from FONDECYT Regular 1181264. FAG, AM and CM acknowledge funding from the Max Planck Society through a Partner Group grant. AM acknowledges financial support from FONDECYT Regular 1181797. FM acknowledges support through
the Program `Rita Levi Montalcini' of the Italian MIUR

\bibliographystyle{apj}
\bibliography{apj-jour,Getal19}

\begin{thebibliography}{53}
\expandafter\ifx\csname natexlab\endcsname\relax\def\natexlab#1{#1}\fi

\bibitem[{{Alfaro} {et~al.}(2001){Alfaro}, {P{\'e}rez}, {Gonz{\'a}lez Delgado},
  {Martos}, \& {Franco}}]{2001ApJ...550..253A}
{Alfaro}, E.~J., {P{\'e}rez}, E., {Gonz{\'a}lez Delgado}, R.~M., {Martos},
  M.~A., \& {Franco}, J. 2001, ApJ, 550, 253

\bibitem[{{Ann} \& {Park}(2006)}]{2006NewA...11..293A}
{Ann}, H.~B., \& {Park}, J.-C. 2006, \na, 11, 293

\bibitem[{{Antoja} {et~al.}(2018){Antoja}, {Helmi}, {Romero-G{\'o}mez}, {Katz},
  {Babusiaux}, {Drimmel}, {Evans}, {Figueras}, {Poggio}, {Reyl{\'e}}, {Robin},
  {Seabroke}, \& {Soubiran}}]{2018Natur.561..360A}
{Antoja}, T., {et~al.} 2018, \nat, 561, 360

\bibitem[{{Bland-Hawthorn} {et~al.}(2018){Bland-Hawthorn}, {Sharma},
  {Tepper-Garcia}, {Binney}, {Freeman}, {Hayden}, {Kos}, {De Silva}, {Lewis},
  {Asplund}, {Buder}, {Casey}, {D'Orazi}, {Duong}, {Lin}, {Lind}, {Martell},
  {Ness}, {Simpson}, {Zucker}, {Zwitter}, {Kafle}, {Quillen}, {Ting}, \&
  {Wyse}}]{2018arXiv180902658B}
{Bland-Hawthorn}, J., {et~al.} 2018, ArXiv e-prints

\bibitem[{{Boselli} {et~al.}(2018){Boselli}, {Fossati}, {Ferrarese},
  {Boissier}, {Consolandi}, {Longobardi}, {Amram}, {Balogh}, {Barmby},
  {Boquien}, {Boulanger}, {Braine}, {Buat}, {Burgarella}, {Combes}, {Contini},
  {Cortese}, {C{\^o}t{\'e}}, {C{\^o}t{\'e}}, {Cuilland re}, {Drissen},
  {Epinat}, {Fumagalli}, {Gallagher}, {Gavazzi}, {Gomez-Lopez}, {Gwyn},
  {Harris}, {Hensler}, {Koribalski}, {Marcelin}, {McConnachie},
  {Miville-Deschenes}, {Navarro}, {Patton}, {Peng}, {Plana}, {Prantzos},
  {Robert}, {Roediger}, {Roehlly}, {Russeil}, {Salome}, {Sanchez-Janssen},
  {Serra}, {Spekkens}, {Sun}, {Taylor}, {Tonnesen}, {Vollmer}, {Willis},
  {Wozniak}, {Burdullis}, {Devost}, {Mahoney}, {Manset}, {Petric}, {Prunet}, \&
  {Withington}}]{2018A&A...614A..56B}
{Boselli}, A., {et~al.} 2018, \aap, 614, A56

\bibitem[{{Canzian}(1993)}]{1993ApJ...Canzian}
{Canzian}, B. 1993, \apj, 414, 487

\bibitem[{{Carpintero} {et~al.}(2014){Carpintero}, {Maffione}, \&
  {Darriba}}]{2014A&C.....5...19C}
{Carpintero}, D.~D., {Maffione}, N., \& {Darriba}, L. 2014, Astronomy and
  Computing, 5, 19

\bibitem[{{Cautun} {et~al.}(2020){Cautun}, {Ben{\'\i}tez-Llambay}, {Deason},
  {Frenk}, {Fattahi}, {G{\'o}mez}, {Grand}, {Oman}, {Navarro}, \&
  {Simpson}}]{2020MNRAS.494.4291C}
{Cautun}, M., {et~al.} 2020, \mnras, 494, 4291

\bibitem[{{Daigle} {et~al.}(2006){Daigle}, {Carignan}, {Hernandez}, {Chemin},
  \& {Amram}}]{2006MNRAS.368.1016D}
{Daigle}, O., {Carignan}, C., {Hernandez}, O., {Chemin}, L., \& {Amram}, P.
  2006, MNRAS, 368, 1016

\bibitem[{{D'Onghia} {et~al.}(2016){D'Onghia}, {Madau}, {Vera-Ciro}, {Quillen},
  \& {Hernquist}}]{2016ApJ...823....4D}
{D'Onghia}, E., {Madau}, P., {Vera-Ciro}, C., {Quillen}, A., \& {Hernquist}, L.
  2016, \apj, 823, 4

\bibitem[{{Epinat} {et~al.}(2008){Epinat}, {Amram}, {Marcelin}, {Balkowski},
  {Daigle}, {Hernandez}, {Chemin}, {Carignan}, {Gach}, \&
  {Balard}}]{2008MNRAS.388..500E}
{Epinat}, B., {et~al.} 2008, MNRAS, 388, 500

\bibitem[{{Fridman} {et~al.}(1998){Fridman}, {Koruzhii}, {Zasov}, {Sil'chenko},
  {Moiseev}, {Burlak}, {Afanas'ev}, {Dodonov}, \&
  {Knapen}}]{1998AstL...24..764F}
{Fridman}, A.~M., {et~al.} 1998, Astronomy Letters, 24, 764

\bibitem[{{Gaia Collaboration} {et~al.}(2018){Gaia Collaboration}, {Katz},
  {Antoja}, {Romero-G{\'o}mez}, {Drimmel}, {Reyl{\'e}}, {Seabroke}, {Soubiran},
  {Babusiaux}, {Di Matteo}, {Figueras}, {Poggio}, {Robin}, {Evans}, \& {DPAC
  co-authors}}]{2018arXiv180409380G}
{Gaia Collaboration} {et~al.} 2018, ArXiv e-prints

\bibitem[{{G{\'o}mez} {et~al.}(2013){G{\'o}mez}, {Minchev}, {O'Shea}, {Beers},
  {Bullock}, \& {Purcell}}]{2013MNRAS.429..159G}
{G{\'o}mez}, F.~A., {Minchev}, I., {O'Shea}, B.~W., {Beers}, T.~C., {Bullock},
  J.~S., \& {Purcell}, C.~W. 2013, MNRAS, 429, 159

\bibitem[{{G{\'o}mez} {et~al.}(2017){G{\'o}mez}, {White}, {Grand}, {Marinacci},
  {Springel}, \& {Pakmor}}]{2017MNRAS.465.3446G}
{G{\'o}mez}, F.~A., {White}, S.~D.~M., {Grand}, R.~J.~J., {Marinacci}, F.,
  {Springel}, V., \& {Pakmor}, R. 2017, MNRAS, 465, 3446

\bibitem[{{G{\'o}mez} {et~al.}(2016){G{\'o}mez}, {White}, {Marinacci},
  {Slater}, {Grand}, {Springel}, \& {Pakmor}}]{2016MNRAS.456.2779G}
{G{\'o}mez}, F.~A., {White}, S.~D.~M., {Marinacci}, F., {Slater}, C.~T.,
  {Grand}, R.~J.~J., {Springel}, V., \& {Pakmor}, R. 2016, MNRAS, 456, 2779

\bibitem[{{G{\'o}mez} {et~al.}(2012){G{\'o}mez}, {Minchev}, {O'Shea}, {Lee},
  {Beers}, {An}, {Bullock}, {Purcell}, \& {Villalobos}}]{2012MNRAS.423.3727G}
{G{\'o}mez}, F.~A., {et~al.} 2012, MNRAS, 423, 3727

\bibitem[{{G{\'o}mez-L{\'o}pez} {et~al.}(2019){G{\'o}mez-L{\'o}pez}, {Amram},
  {Epinat}, {Boselli}, {Rosado}, {Marcelin}, {Boissier}, {Gach},
  {S{\'a}nchez-Cruces}, \& {Sardaneta}}]{2019A&A...631A..71G}
{G{\'o}mez-L{\'o}pez}, J.~A., {et~al.} 2019, \aap, 631, A71

\bibitem[{{Grand} {et~al.}(2016{\natexlab{a}}){Grand}, {Springel}, {G{\'o}mez},
  {Marinacci}, {Pakmor}, {Campbell}, \& {Jenkins}}]{2016MNRAS.459..199G}
{Grand}, R. J.~J., {Springel}, V., {G{\'o}mez}, F.~A., {Marinacci}, F.,
  {Pakmor}, R., {Campbell}, D. J.~R., \& {Jenkins}, A. 2016{\natexlab{a}},
  \mnras, 459, 199

\bibitem[{{Grand} {et~al.}(2016{\natexlab{b}}){Grand}, {Springel}, {Kawata},
  {Minchev}, {S{\'a}nchez-Bl{\'a}zquez}, {G{\'o}mez}, {Marinacci}, {Pakmor}, \&
  {Campbell}}]{2016MNRAS.460L..94G}
{Grand}, R.~J.~J., {et~al.} 2016{\natexlab{b}}, MNRAS, 460, L94

\bibitem[{{Grand} {et~al.}(2017){Grand}, {G{\'o}mez}, {Marinacci}, {Pakmor},
  {Springel}, {Campbell}, {Frenk}, {Jenkins}, \& {White}}]{2017MNRAS.467..179G}
---. 2017, MNRAS, 467, 179

\bibitem[{{Grand} {et~al.}(2019){Grand}, {van de Voort}, {Zjupa}, {Fragkoudi},
  {G{\'o}mez}, {Kauffmann}, {Marinacci}, {Pakmor}, {Springel}, \&
  {White}}]{2019MNRAS.490.4786G}
{Grand}, R. J.~J., {et~al.} 2019, \mnras, 490, 4786

\bibitem[{{Guglielmo} {et~al.}(2018){Guglielmo}, {Lane}, {Conn}, {Ho}, {Ibata},
  \& {Lewis}}]{2018MNRAS.474.4584G}
{Guglielmo}, M., {Lane}, R.~R., {Conn}, B.~C., {Ho}, A.~Y.~Q., {Ibata}, R.~A.,
  \& {Lewis}, G.~F. 2018, MNRAS, 474, 4584

\bibitem[{{Ho} {et~al.}(2011){Ho}, {Li}, {Barth}, {Seigar}, \&
  {Peng}}]{2011ApJS..197...21H}
{Ho}, L.~C., {Li}, Z.-Y., {Barth}, A.~J., {Seigar}, M.~S., \& {Peng}, C.~Y.
  2011, \apjs, 197, 21

\bibitem[{{Laporte} {et~al.}(2018{\natexlab{a}}){Laporte}, {G{\'o}mez},
  {Besla}, {Johnston}, \& {Garavito-Camargo}}]{2018MNRAS.473.1218L}
{Laporte}, C.~F.~P., {G{\'o}mez}, F.~A., {Besla}, G., {Johnston}, K.~V., \&
  {Garavito-Camargo}, N. 2018{\natexlab{a}}, MNRAS, 473, 1218

\bibitem[{{Laporte} {et~al.}(2018{\natexlab{b}}){Laporte}, {Johnston},
  {G{\'o}mez}, {Garavito-Camargo}, \& {Besla}}]{2018MNRAS.481..286L}
{Laporte}, C.~F.~P., {Johnston}, K.~V., {G{\'o}mez}, F.~A., {Garavito-Camargo},
  N., \& {Besla}, G. 2018{\natexlab{b}}, \mnras, 481, 286

\bibitem[{{Laporte} {et~al.}(2019){Laporte}, {Minchev}, {Johnston}, \&
  {G{\'o}mez}}]{2019MNRAS.485.3134L}
{Laporte}, C. F.~P., {Minchev}, I., {Johnston}, K.~V., \& {G{\'o}mez}, F.~A.
  2019, \mnras, 485, 3134

\bibitem[{{Liu} {et~al.}(2017){Liu}, {Xu}, {Wan}, {Wang}, {Carlin}, {Deng},
  {Newberg}, {Cao}, {Hou}, {Wang}, \& {Zhang}}]{2017RAA....17...96L}
{Liu}, C., {et~al.} 2017, Research in Astronomy and Astrophysics, 17, 096

\bibitem[{{Marinacci} {et~al.}(2014){Marinacci}, {Pakmor}, \&
  {Springel}}]{2014MNRAS.437.1750M}
{Marinacci}, F., {Pakmor}, R., \& {Springel}, V. 2014, MNRAS, 437, 1750

\bibitem[{{McMillan}(2017)}]{2017MNRAS.465...76M}
{McMillan}, P.~J. 2017, MNRAS, 465, 76

\bibitem[{{Miki} \& {Umemura}(2018)}]{2018MNRAS.475.2269M}
{Miki}, Y., \& {Umemura}, M. 2018, \mnras, 475, 2269

\bibitem[{{Miyamoto} \& {Nagai}(1975)}]{1975PASJ...27..533M}
{Miyamoto}, M., \& {Nagai}, R. 1975, \pasj, 27, 533

\bibitem[{{Monari} {et~al.}(2016){Monari}, {Famaey}, {Siebert}, {Grand },
  {Kawata}, \& {Boily}}]{2016MNRAS.461.3835M}
{Monari}, G., {Famaey}, B., {Siebert}, A., {Grand }, R. J.~J., {Kawata}, D., \&
  {Boily}, C. 2016, \mnras, 461, 3835

\bibitem[{{Mu{\~n}oz-Elgueta} {et~al.}(2018){Mu{\~n}oz-Elgueta},
  {Torres-Flores}, {Amram}, {Hernandez-Jimenez}, {Urrutia-Viscarra}, {Mendes de
  Oliveira}, \& {G{\'o}mez-L{\'o}pez}}]{2018MNRAS.480.3257M}
{Mu{\~n}oz-Elgueta}, N., {Torres-Flores}, S., {Amram}, P., {Hernandez-Jimenez},
  J.~A., {Urrutia-Viscarra}, F., {Mendes de Oliveira}, C., \&
  {G{\'o}mez-L{\'o}pez}, J.~A. 2018, \mnras, 480, 3257

\bibitem[{{Newberg} {et~al.}(2002){Newberg}, {Yanny}, {Rockosi}, {Grebel},
  {Rix}, {Brinkmann}, {Csabai}, {Hennessy}, {Hindsley}, {Ibata}, {Ivezi{\'c}},
  {Lamb}, {Nash}, {Odenkirchen}, {Rave}, {Schneider}, {Smith}, {Stolte}, \&
  {York}}]{new02}
{Newberg}, H.~J., {et~al.} 2002, ApJ, 569, 245

\bibitem[{{Pogge} \& {Martini}(2002)}]{2002ApJ...569..624P}
{Pogge}, R.~W., \& {Martini}, P. 2002, \apj, 569, 624

\bibitem[{{Pranav} \& {Jog}(2010)}]{2010MNRAS.406..576P}
{Pranav}, P., \& {Jog}, C.~J. 2010, \mnras, 406, 576

\bibitem[{{Price-Whelan} {et~al.}(2015){Price-Whelan}, {Johnston}, {Sheffield},
  {Laporte}, \& {Sesar}}]{2015MNRAS.452..676P}
{Price-Whelan}, A.~M., {Johnston}, K.~V., {Sheffield}, A.~A., {Laporte},
  C.~F.~P., \& {Sesar}, B. 2015, MNRAS, 452, 676

\bibitem[{{Purcell} {et~al.}(2011){Purcell}, {Bullock}, {Tollerud}, {Rocha}, \&
  {Chakrabarti}}]{2011Natur.477..301P}
{Purcell}, C.~W., {Bullock}, J.~S., {Tollerud}, E.~J., {Rocha}, M., \&
  {Chakrabarti}, S. 2011, \nat, 477, 301

\bibitem[{{Reshetnikov} \& {Combes}(1998)}]{1998A&A...337....9R}
{Reshetnikov}, V., \& {Combes}, F. 1998, \aap, 337, 9

\bibitem[{{S{\'a}nchez-Gil} {et~al.}(2015){S{\'a}nchez-Gil}, {Alfaro}, \&
  {P{\'e}rez}}]{2015MNRAS.454.3376S}
{S{\'a}nchez-Gil}, M.~C., {Alfaro}, E.~J., \& {P{\'e}rez}, E. 2015, MNRAS, 454,
  3376

\bibitem[{{Schaye} {et~al.}(2015){Schaye}, {Crain}, {Bower}, {Furlong},
  {Schaller}, {Theuns}, {Dalla Vecchia}, {Frenk}, {McCarthy}, {Helly},
  {Jenkins}, {Rosas-Guevara}, {White}, {Baes}, {Booth}, {Camps}, {Navarro},
  {Qu}, {Rahmati}, {Sawala}, {Thomas}, \& {Trayford}}]{2015MNRAS.446..521S}
{Schaye}, J., {et~al.} 2015, MNRAS, 446, 521

\bibitem[{{Sheffield} {et~al.}(2018){Sheffield}, {Price-Whelan}, {Tzanidakis},
  {Johnston}, {Laporte}, \& {Sesar}}]{2018ApJ...854...47S}
{Sheffield}, A.~A., {Price-Whelan}, A.~M., {Tzanidakis}, A., {Johnston}, K.~V.,
  {Laporte}, C.~F.~P., \& {Sesar}, B. 2018, ApJ, 854, 47

\bibitem[{{Shen} \& {Sellwood}(2006)}]{2006MNRAS.370....2S}
{Shen}, J., \& {Sellwood}, J.~A. 2006, \mnras, 370, 2

\bibitem[{{Siebert} {et~al.}(2012){Siebert}, {Famaey}, {Binney}, {Burnett},
  {Faure}, {Minchev}, {Williams}, {Bienaym{\'e}}, {Bland-Hawthorn}, {Boeche},
  {Gibson}, {Grebel}, {Helmi}, {Just}, {Munari}, {Navarro}, {Parker}, {Reid},
  {Seabroke}, {Siviero}, {Steinmetz}, \& {Zwitter}}]{2012MNRAS.425.2335S}
{Siebert}, A., {et~al.} 2012, MNRAS, 425, 2335

\bibitem[{{Slater} {et~al.}(2014){Slater}, {Bell}, {Schlafly}, {Morganson},
  {Martin}, {Rix}, {Pe{\~n}arrubia}, {Bernard}, {Ferguson}, {Martinez-Delgado},
  {Wyse}, {Burgett}, {Chambers}, {Draper}, {Hodapp}, {Kaiser}, {Magnier},
  {Metcalfe}, {Price}, {Tonry}, {Wainscoat}, \& {Waters}}]{2014ApJ...791....9S}
{Slater}, C.~T., {et~al.} 2014, ApJ, 791, 9

\bibitem[{{Smith} {et~al.}(2015){Smith}, {Flynn}, {Candlish}, {Fellhauer}, \&
  {Gibson}}]{2015MNRAS.448.2934S}
{Smith}, R., {Flynn}, C., {Candlish}, G.~N., {Fellhauer}, M., \& {Gibson},
  B.~K. 2015, \mnras, 448, 2934

\bibitem[{{Springel}(2010)}]{2010MNRAS.401..791S}
{Springel}, V. 2010, MNRAS, 401, 791

\bibitem[{{Torres-Flores} {et~al.}(2014){Torres-Flores}, {Amram}, {Mendes de
  Oliveira}, {Plana}, {Balkowski}, {Marcelin}, \&
  {Olave-Rojas}}]{2014MNRAS.442.2188T}
{Torres-Flores}, S., {Amram}, P., {Mendes de Oliveira}, C., {Plana}, H.,
  {Balkowski}, C., {Marcelin}, M., \& {Olave-Rojas}, D. 2014, MNRAS, 442, 2188

\bibitem[{{Vesperini} \& {Weinberg}(2000)}]{2000ApJ...534..598V}
{Vesperini}, E., \& {Weinberg}, M.~D. 2000, ApJ, 534, 598

\bibitem[{{Widrow} {et~al.}(2012){Widrow}, {Gardner}, {Yanny}, {Dodelson}, \&
  {Chen}}]{2012ApJ...750L..41W}
{Widrow}, L.~M., {Gardner}, S., {Yanny}, B., {Dodelson}, S., \& {Chen}, H.-Y.
  2012, ApJL, 750, L41

\bibitem[{{Xu} {et~al.}(2015){Xu}, {Newberg}, {Carlin}, {Liu}, {Deng}, {Li},
  {Sch{\"o}nrich}, \& {Yanny}}]{2015ApJ...801..105X}
{Xu}, Y., {Newberg}, H.~J., {Carlin}, J.~L., {Liu}, C., {Deng}, L., {Li}, J.,
  {Sch{\"o}nrich}, R., \& {Yanny}, B. 2015, ApJ, 801, 105

\bibitem[{{Yanny} {et~al.}(2003){Yanny}, {Newberg}, {Grebel}, {Kent},
  {Odenkirchen}, {Rockosi}, {Schlegel}, {Subbarao}, {Brinkmann}, {Fukugita},
  {Ivezic}, {Lamb}, {Schneider}, \& {York}}]{yanny03}
{Yanny}, B., {et~al.} 2003, ApJ, 588, 824

\end{thebibliography}




\appendix
\label{app:1}
\section{M/L ratios across a simulated disc}

In order to test whether the assumption of a  mass-to-light (M/L) ratio across the disc is reasonable, we 
computed the M/L ratio map for the galactic disc of the Au25 model. The map is shown in Figure 12. As in VV304a observations, we have focused on the r'-band. The reason behind this choice is that this band is a better tracer of the overall mass distribution than bluer photometric bands. As we can see from this Figure, our model indicates an average M/L of the order 1 throughout the disc, with values that can vary between 0.2 to 2.75. This  indicates that assuming a constant M/L in the analysis shown in Section 2.3 is a reasonable approximation in this case. Note that our goal in that Section is to estimate the strength of the spiral structure in terms of its overdensity pattern. Our study suggests that the spiral structure of VV304a is $\sim$ 30 per cent denser than the background density of the overall disc. However, in our analytic model, we have gone as far as considering spiral structures that are 1000\% denser than the background density. Thus, we expect small departures from a constant M/L to be covered by our most extreme models.

\begin{figure}
    \centering
    \includegraphics[width=85mm,clip,angle=0]{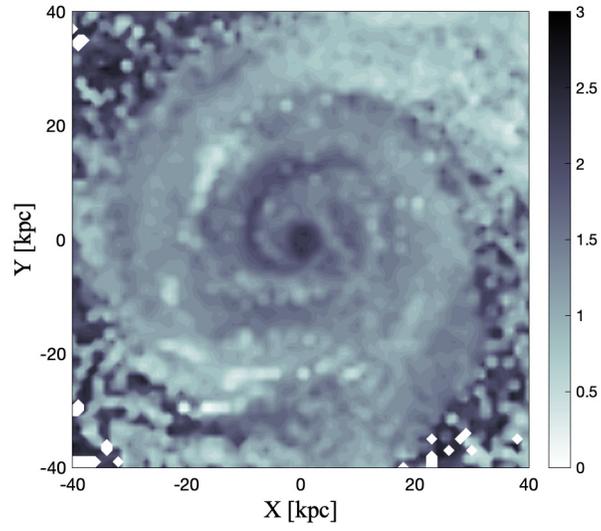}
    \caption{The color coding indicates the values of the simulated mass-to-light ratio in the r'-band, M/L, obtained across the disc of the cosmological simulation Au25. Values vary between M/L $\sim 0.2$ and 2.75.}
    \label{fig:5}
\end{figure}



\end{document}